\documentclass[apj]{emulateapj}

\slugcomment{}

\shorttitle{}
\shortauthors{}

\usepackage{color}
\usepackage{amsmath}
\usepackage{booktabs}
\usepackage{enumerate}
\bibpunct{(}{)}{;}{a}{}{,}

\begin{document}


\title{Probing Episodic Accretion in Very Low Luminosity Objects}

\author{Tien-Hao Hsieh$^{1,2}$, Nadia M. Murillo$^{3}$, Arnaud Belloche$^{4}$, Naomi Hirano$^{1}$, Catherine Walsh$^{5}$, Ewine F. van Dishoeck$^{3,6}$, Shih-Ping Lai$^{1,2}$}
\affil{$^{1}$Institute of Astronomy and Astrophysics, Academia Sinica, P.O. Box 23-141, Taipei 106, Taiwan}
\affil{$^{2}$Institute of Astronomy, National Tsing Hua University (NTHU), Hsinchu 30013, Taiwan}
\affil{$^{3}$Leiden Observatory, Leiden University, P.O. Box 9513, 2300 RA, Leiden, the Netherlands}
\affil{$^{4}$Max-Planck-Institut f\"{u}r Radioastronomie, Auf dem H\"{u}gel 69, 53121 Bonn, Germany}
\affil{$^{5}$School of Physics and Astronomy, University of Leeds, Leeds LS2 9JT, UK}
\affil{$^{6}$Max-Planck-Institut f\"{u}r extraterrestrische Physik, Giessenbachstra{\ss}e 1, 85748, Garching bei M\"{u}nchen, Germany}

\email{thhsieh@asiaa.sinica.edu.tw}

\begin{abstract}
Episodic accretion has been proposed as a solution to the long-standing luminosity problem in star formation; however, the process remains poorly understood.
We present observations of line emission from N$_2$H$^+$ and CO isotopologues using the Atacama Large Millimeter/submillimeter Array 
(ALMA) in the envelopes of eight Very Low Luminosity Objects (VeLLOs).
In five of the sources the spatial distribution of emission from N$_2$H$^+$ and 
CO isotopologues shows a clear anti-correlation. It is proposed that this is tracing the CO snow line in the envelopes: 
N$_2$H$^+$ emission is depleted toward the center of these sources in contrast to the CO isotopologue emission which exhibits a peak. 
The positions of the CO snow lines traced by the N$_2$H$^+$ emission are located at much larger radii than those calculated using the current luminosities of the central sources. 
This implies that these five sources have experienced a recent accretion burst because the CO snow line would have been pushed outwards during the burst due to the 
increased luminosity of the central star.
The N$_2$H$^+$ and CO isotopologue emission from DCE161, one of the other three sources, 
is most likely tracing a transition disk at a later evolutionary stage. 
Excluding DCE161, five out of seven sources (i.e., $\sim$70\%) show signatures of a recent accretion burst. This fraction is larger than that of the Class 0/I sources studied by 
\citet{jo15} and \citet{fr16} suggesting that the interval between accretion episodes in VeLLOs is shorter than that in Class 0/I sources.
\end{abstract}



\section{INTRODUCTION}
The long-standing luminosity problem was first noted by \citet{ke90}.
The bolometric luminosity, dominated by the accretion luminosity $L_{\rm acc}$ \citep{ha96}, is predicted to be a few tens of $L_{\odot}$ \citep{of11,du14} assuming a typical mass 
accretion rate of $\sim2\times10^{-6}~M_\odot$ yr$^{-1}$ \citep{sh77,te84,mc07}.
On the other hand, bolometric luminosities derived from surveys covering large samples of young stellar objects 
(YSOs) are found to be much lower than the predicted luminosity \citep{ev09,en09,kr12,du13}.
\citet{ke90} first proposed the episodic accretion process to explain this discrepancy and this
is now considered to be the most plausible explanation. 
This mechanism proposes that a protostellar system is in a quiescent accretion phase most 
of the time with occasional accretion bursts that deliver material onto the central protostar.
This process predicts a low protostellar luminosity for the majority of the time whilst also enabling the central source to acquire sufficient material to form a star.

A number of theories have been proposed to explain the origin of episodic accretion \citep{au14}. Of these, the favoured origin is accretion bursts driven by disk instability.
Evidence for the presence of unstable disks comes from recent high-resolution 
near-infrared images for four sources undergoing accretion bursts \citep{li16}.
Disk instabilities may arise due to several mechanisms including thermal instability \citep{li85,be94,ba10}, 
gravitational instability \citep{vo05,bo08}, or a combination of both \citep{ar01,zh09}.
In addition, stellar (or planetary) encounters can be a possible trigger of protostellar 
episodic accretion \citep{cl96,lo04,fo10}.

\tabletypesize{\scriptsize}
\tabcolsep=0.08cm
\begin{deluxetable*}{cccccccccccc}
\tabletypesize{\tiny}
\tablecaption{Target list}
\tablehead{ 
\colhead{Source}
& \colhead{Other name}
& \colhead{Region\tablenotemark{a} }
& \colhead{R.A.}
& \colhead{Dec}
& \colhead{$L_\textmd{int}$}
& \colhead{$L_\textmd{bol}$}
& \colhead{$T_\textmd{bol}$} 
& \colhead{distance}
& \colhead{N$_2$D$^+$/N$_2$H$^+$\tablenotemark{b}}
& \colhead{Opening Angle.\tablenotemark{c}}
& \colhead{Ref.}\\
\colhead{}		
& \colhead{}		
& \colhead{}
& \colhead{}	
& \colhead{}
& \colhead{(L$_{\odot}$)}	
& \colhead{(L$_{\odot}$)}			
& \colhead{(K)}	
& \colhead{(pc)}		
& \colhead{}	
& \colhead{(degree)}
}
\startdata 
DCE018	& 			& DC 3272+18 &15:42:16.99	& -52:48:02.2	& 0.04	& $0.06\pm0.01$	& $105\pm3$	& 250	& ...				& ...  & -\\
DCE024	& CB130-1-IRS1& CB 130-3	& 18:16:16.39 & -02:32:37.7	 & 0.07	& $0.20\pm0.04$	& $55\pm10$ 	& 270	& ...				& 15$\pm$2.5 & 1\\      
DCE031	& L673-7		& L673-7		& 19:21:34.82 & +11:21:23.4 	& 0.04	& $0.09\pm0.03$	& $24\pm6$	& 300	& 0.040$\pm$0.006 & N & 2\\
DCE064	& 			& Perseus		& 03:28:32.57 & +31:11:05.3 	& 0.03	& $0.20\pm0.05$	& $65\pm12$	& 250	& 0.018$\pm$0.006	& 55$\pm$2.5 & -\\
DCE065	& 			& Perseus		& 03:28:39.10 & +31:06:01.8 	& 0.02	& $0.22\pm0.06$	& $29\pm3$	& 250	& 0.087$\pm$0.011	& N & 3\\ 
DCE081	&	 		& Perseus		&03:30:32.69	& 30:26:26.5	& 0.06	& $0.18\pm0.04$	& $33\pm4$	& 250	& 0.028$\pm$0.002 	& N & -\\
DCE161	& 			& Lupus IV	&16:01:15.55	& -41:52:35.4	& 0.08	& $\leq0.11$		& $\leq126$	& 150	& ...				& ... & -\\
DCE185	& IRAS 16253-2429	& Ophiuchus& 16:28:21.60 & -24:36:23.4 & 0.09	& $0.45\pm0.08$	& $30\pm2$ 	& 125	& 0.064$\pm$0.005	& $<$35 & 4,5,6\\
\enddata
\tablecomments{
References:
(1) \citealp{ki11};
(2) \citealp{du10a};
(3) \citealp{hu10};
(4) \citealp{ye15};
(5) \citealp{hs16};
(6) \citealp{ye17}}
\tablenotetext{a}{Regions are defined the same as defined by the c2d team \citep{ev09}.}
\tablenotetext{b}{N$_2$D$^+$/N$_2$H$^+$ abundance ratios from \citet{hs15}.}
\tablenotetext{c}{The outflow opening angles were measured through near-infrared scattered light by \citet{hs17}. 
``N'' stands for sources that were observed but not detected in this infrared study.}
\label{tab:targets}
\end{deluxetable*}

Theoretical models predict a number of characteristics of the episodic accretion process 
which  require observational confirmation.
\citet{vo05,vo10} modeled the collapse of a rotating cloud and found that dense clumps formed through disk fragmentation can fall onto the central star and trigger an accretion burst.
In this scenario, the episodic accretion process is more prone to occur at the Class I stage when the 
disk is sufficiently massive to fragment \citep{vo13,vo15}.
Besides, in a more continuous accretion process, radiative feedback can suppress fragmentation 
by heating the cloud core above 100~K \citep{of09,yi12,yi15,kr14}.
\citet{st12} proposed that episodic accretion can moderate the effect of radiative feedback provided that there is sufficient time for a disk to cool and fragment in the quiescent phase.
Further, this allows the formation of low-mass stars, brown dwarfs, and planetary-mass objects through 
fragmentation of the protostellar disk.  
This mechanism in turn highlights that the interval between accretion episodes may be
critical for determining the low-mass end of the initial mass function \citep{st11,me16}.

Recent observations have provided direct or indirect evidence for the occurrence of 
episodic accretion in low-mass protostars.
Young stellar objects undergoing luminosity outbursts (e.g., \citealp[FU Orionis and EX Orionis events;][]{he66,he77}) 
are considered to be direct evidence for episodic accretion \citep{au14}.
Direct detection of an accretion burst is difficult due to the relatively long time interval between bursts 
(\citealp[$\sim5\times10^{3}-5\times10^{4}$ yr;][]{sc13}).  
Hence, only a few cases have been reported to date 
(\citealp[V1647 Ori:][]{ab04,an04,ac07,fe07,as09}, \citealp[OO Serpentis:][]{ko07}, 
\citealp[\lbrack CTF93\rbrack 216-2:][]{ca11}, \citealp[VSX J205126.1:][]{co11,ko11}, \citealp[HOPS 383:][]{sa15}).
However, all sources except HOPS 383 \citep{sa15} are at the very late Class I or later stage, 
near the end of the embedded phase \citep{co11,ko11}. 
Luminosity variations in the earlier embedded phase will be probed by an ongoing
sub-millimeter survey that monitors 182 Class 0/I objects over 3.5 yr \citep{he17}.

Apart from the direct detection of luminosity variations, chemical signatures provide an excellent 
way to trace the episodic accretion process. This is because the chemical composition of the gas is sensitive to temperature changes 
\citep{ki11,ki12,vi12,vi15,ta16}.
The location of snow lines (or sublimation radii) of molecules is thus a powerful indicator of the 
thermal history of protostellar envelopes.
Using the water snow line tracers, H$^{13}$CO$^+$ and CH$_3$OH, \citet{jo13} found that 
IRAS 15398--3359 has likely experienced a recent accretion burst 
because the snow line is located at a radius larger than that expected from the current luminosity.
The outwards shift of the water snow line in IRAS 15398--3359 is further confirmed by HDO 
observations \citep{bj16}.
Similarly, the CO snow line can be used to probe episodic accretion \citep{jo15,an16,fr17}.
The radii of several CO snow lines reported by \citet{jo15} and \citet{fr17} are larger than those 
predicted from the current luminosities. 
They concluded that $20-50\%$ of their sample sources have experienced recent accretion bursts. 
Assuming a time scale of $\sim$10,000 yr for CO to refreeze out \citep{vi15}, the interval 
between accretion bursts was estimated to be $2-5\times10^{4}$ yr. This is comparable to the value derived from the luminosity monitoring of 4000 YSOs by 
\citet{sc13}. 
On the other hand, the CO snow lines in four Class 0 protostars studied by 
\citet{an16} did not reside at larger radii than the expected values.

Very Low Luminosity Objects (VeLLOs) were first discovered by the {\em Spitzer} Space Telescope \citep{yo04}  
and are defined as YSOs with internal luminosities $L_{\rm int}<0.1~L_\odot$ \citep{di07}.
Given their low internal luminosities the discovery of VeLLOs exacerbates the luminosity problem. 
One explanation is that VeLLOs are in a quiescent phase of the episodic accretion process.
This hypothesis is supported by several studies (\citealp[L673-7:][]{du10b}, \citealp[L1521F:][]{ta13}) that found that the averaged mass accretion rates derived from molecular outflows from these YSOs are a few times higher than that inferred from their internal luminosity.
Furthermore, the low internal luminosity also suggests that VeLLOs are either young Class 0 sources 
(\citealp[IRAM 04191:][]{an99,be02,du06}, \citealp[L1521F:][]{bo06,ta13}, \citealp[Cha-MMS1:][]{be06,ts13,va14}, 
\citealp[IRAS16253:][]{hs15,hs17}) or extremely low-mass protostars or even proto-brown dwarf candidates 
(\citealp[L328:][]{le09,le13}, \citealp[L1148:][]{ka11}, \citealp[IC 348-SMM2E:][]{pa14}). 
In any case, VeLLOs are unlikely to have a massive disk prone to fragmentation.

\tabletypesize{\scriptsize}
\tablewidth{0pt}
\tabcolsep=0.08cm
\begin{deluxetable*}{ccccccccccccccc}
\tabletypesize{\tiny}
\tablecaption{Parameters of molecular integrated intensity maps and of continuum maps}
\tablehead{ 
\colhead{}
& \colhead{}
& \multicolumn{2}{c}{N$_2$H$^+$ (1--0)}
& \colhead{}
& \multicolumn{2}{c}{$^{13}$CO (1--0)}
& \multicolumn{2}{c}{C$^{18}$O (1--0)}
& \multicolumn{2}{c}{C$^{17}$O (1--0)}
& \multicolumn{3}{c}{Continuum}
\\ 
\cmidrule(lr){3-4} \cmidrule(lr){6-7} \cmidrule(lr){8-9} \cmidrule(lr){10-11} \cmidrule(lr){12-14}
\colhead{Source}		
& \colhead{Vel. range}		
& \colhead{beam}
& \colhead{rms}	
& \colhead{Vel. range}	
& \colhead{beam}
& \colhead{rms}	
& \colhead{beam}
& \colhead{rms}
& \colhead{beam}
& \colhead{rms}
& \colhead{beam}
& \colhead{rms}
& \colhead{peak (S/N)}
\\
\colhead{}		
& \colhead{(km s$^{-1}$)}		
& \colhead{(arcsec)}
& \colhead{}
& \colhead{(km s$^{-1}$)}	
& \colhead{(arcsec)}
& \colhead{}
& \colhead{(arcsec)}
& \colhead{}
& \colhead{(arcsec)}
& \colhead{}
& \colhead{(arcsec)}
& \colhead{($\mu$Jy beam$^{-1}$)}
& \colhead{($\mu$Jy beam$^{-1}$)}
}
\startdata 
DCE018	& -0.52--0.25	& 2.13$\times$1.42	& 9.5	& -0.45---0.05	& 2.27$\times$1.84	& 2.7	& 2.29$\times$1.89	& 2.6	& 2.22$\times$1.89	& 5.1	& 2.08$\times$1.53	& 46	& 2003 (43.1) \\
DCE024	& 7.12--8.02	& 3.14$\times$1.67	& 15.3	& 7.25--7.75	& 2.08$\times$1.73	& 3.6	& 2.08$\times$1.77	& 3.1	& 1.99$\times$1.77	& 6.1	& 2.26$\times$1.64	& 50	& 1839 (36.1) \\
DCE031	& 6.54--8.08	& 2.75$\times$1.67	& 19.5	& 6.90--7.50	& 1.78$\times$1.58	& 3.7	& 1.83$\times$1.65	& 3.5	& 1.77$\times$1.65	& 7.1	& 1.88$\times$1.66	& 53	& 1439 (26.7) \\
DCE064	& 6.65--7.55	& 3.17$\times$2.03	& 19.0	& 7.35--7.85	& 2.94$\times$1.79	& 3.9	& 2.94$\times$1.80	& 4.2	& 2.92$\times$1.80	& 8.2	& 2.76$\times$1.82	& 60	& 324 (5.4) \\
DCE065	& 6.73--7.29	& 3.17$\times$2.03	& 9.8	& 6.25--6.95	& 2.94$\times$1.79	& 6.1	& 2.94$\times$1.80	& 5.6	& 2.92$\times$1.80	& 9.9	& 2.75$\times$1.83	& 55	& 233 (4.3) \\
DCE081	& 5.91--6.41	& 3.12$\times$2.03	& 14.3	& 6.30--6.90	& 2.88$\times$1.79	& 4.7	& 2.88$\times$1.80	& 4.9	& 2.87$\times$1.80	& 8.5	& 2.71$\times$1.82	& 61	& 212 (3.4) \\
DCE161	& 3.31--4.25	& 2.07$\times$1.37	& 10.5	& 3.90--4.50	& 1.70$\times$1.28	& 3.6	& 1.69$\times$1.22	& 3.4	& 1.64$\times$1.22	& 5.8	& 1.73$\times$1.20	& 42	& 2760 (65.7) \\
DCE185	& 3.60--4.29	& 2.71$\times$1.99	& 6.5	& 3.70--4.30	& 1.66$\times$1.16	& 3.6	& 1.66$\times$1.17	& 3.7	& 1.62$\times$1.17	& 7.1	& 2.13$\times$1.52	& 47	& 2320 (48.6)
\enddata
\tablecomments{For N$_2$H$^+$ (1--0) and C$^{17}$O (1--0), all the seven hyperfine components are integrated over 
the velocity ranges (Col. 2 and 4, respectively) in order to increase the sensitivity.}
\tablecomments{Col. (2) and Col. (5) indicate the integrated velocity ranges for the N$_2$H$^+$ map and the three 
CO isotopologue maps.}
\tablecomments{The rms noise level of the line emissions are in unit of mJy beam$^{-1}$ km s$^{-1}$}
\label{tab:obs}
\end{deluxetable*}

In this paper we present ALMA observations of emission from N$_2$H$^+$ and CO isotopologues towards 
eight VeLLOs. The aims are to search for evidence of episodic accretion using the position of the
CO snow line traced by C$^{18}$O/$^{13}$CO and N$_2$H$^+$ emission.
The sample of VeLLOs and observations are described in Section~\ref{sec:obs}. 
The observational results of both continuum and molecular line emission are given in Section~\ref{sec:results}. The analysis of molecular abundances and modeling of the observations are detailed in Section~\ref{sec:analysis}. 
Finally, the discussion and a summary of the results are given in Sections~\ref{sec:diss} 
and \ref{sec:sum}, respectively.

\section{OBSERVATIONS}
\label{sec:obs}
\subsection{Sample - VeLLOs}
We selected eight VeLLOs (Table \ref{tab:targets}) out of the 15 VeLLOs identified by \citet{du08} based on data 
from the ``{\em Spitzer} Legacy Project: From Molecular Cores to Planet Forming Disks'' (\citealp[c2d,][]{ev03,ev09}).
Sources in our sample are designated with the initials of the first three authors followed by the source number 
in \citet{du08}, e.g., DCE185.
The seven VeLLOs that have been excluded from the full sample can be grouped into 
(1) identified as a galaxy: DCE145 \citep{hs17}; (2) well studied sources: DCE001 
(\citealp[IRAM 04191:][]{an99,be02,be04}), 
DCE038 (\citealp[L1014:][]{cr05,bo05,hu06}), DCE025 (\citealp[L328:][]{le09,le13}), 
and DCE004 (\citealp[L1521F:][]{cr04,bo06,ta13}); 
(3) not observable by ALMA: DCE032 (\citealp[L1148-IRS:][]{ka11}); and (4) at the late 
Class I stage: DCE181 ($T_{\rm bol}=429$ K).
Our eight targets are thus deeply embedded objects ($T_{\rm bol}=24-126$ K) which have not yet been studied in detail.  This is especially true for the two southernmost objects, DCE161 and DCE018.
The distances to the selected targets range from 125 to 300 pc which is sufficiently close to allow us to study 
their cloud core properties in detail.

Some properties of our selected VeLLOs have been reported in the literature.
Among the eight targets, five were observed in a previous single dish survey measuring
N$_2$D$^+$/N$_2$H$^+$ ratios \citep{hs15}, 
and six were included in a previous infrared survey for outflows \citep{hs17}. 
The high deuterium fractionations and small outflow opening angles (Table \ref{tab:targets}) suggest that 
these sources are at a very early evolutionary stage.  DCE064 may be slightly more evolved than the others.
Furthermore, DCE031 and DCE185 are suggested to be undergoing episodic accretion. Using CO outflow observations of DCE031, \citet{du10a} found an average $L_{\rm acc}$ much larger 
than the current $L_{\rm int}$, which indicates a luminosity variation as well as an episodic mass accretion process. 
\citet{hs16} suggested that DCE185 has experienced a recent accretion burst by comparing the position of the 
CO snow line from C$^{18}$O ($2-1$) emission with that expected from the current luminosity.

\subsection{ALMA Observations - N$_2$H$^+$ and 105 GHz continuum}
\label{sec:obs_n2hp}
We observed N$_2$H$^+$ (1--0) and dust continuum emission at 105 GHz simultaneously toward the 
eight VeLLOs from March to May 2016 with ALMA (Cycle 3 project 2015.1.01576.S).
The ALMA configurations were C36-1, C36-2, or C36-3, and the corresponding baselines 
ranged from $\approx 4$ to 100 k$\lambda$ (C36-1) and from $\approx 5$ to 209 k$\lambda$ (C36-3).
The spatial resolution, with natural weighting, is from 1\farcs5 to 3\farcs0 depending on 
the source declination and observing time (Table \ref{tab:obs}).
The channel width of the N$_2$H$^+$ (1--0) observations was 15.259 kHz (0.049 km s$^{-1}$) 
and was binned to 0.05 km s$^{-1}$ in the output maps.
We note that the spectral resolution of the data is 0.098 km s$^{-1}$ because of the 
default ALMA Hanning smoothing.
The continuum spectral window has a bandwidth of 2~GHz at a central frequency of 105 GHz. To enhance the sensitivity, we combined these data with the continuum window at 110 GHz 
(see section \ref{sec:obs110}).

\subsection{ALMA Observations - CO isotopologues and 110 GHz continuum}
\label{sec:obs110}
We used ALMA (in the same project as described in section \ref{sec:obs_n2hp}) 
to simultaneously observe C$^{18}$O (1--0), $^{13}$CO (1--0), C$^{17}$O (1--0), and dust continuum 
at 110~GHz toward the eight VeLLOs from March to May 2016 with the C36-1, C36-2, or C36-3 configurations.
The channel widths were 61.035 kHz ($\sim0.17$ km s$^{-1}$) for $^{13}$CO (1--0) 
and 30.518 kHz (0.08 km s$^{-1}$) for both C$^{18}$O (1--0) and C$^{17}$O (1--0).
These data were later binned to 0.2 km s$^{-1}$ and 0.1 km s$^{-1}$ in the final channel maps, respectively.
The continuum spectral window has a bandwidth of 2~GHz with a central frequency of 110~GHz 
and the data were later combined with the continuum window at 105~GHz to increase 
the sensitivity using the CLEAN algorithm in CASA.
Table \ref{tab:obs} lists the resulting rms noise levels of the continuum maps and the 
integrated line intensity maps.

\begin{figure*}
\centering 
\includegraphics[width=\textwidth]{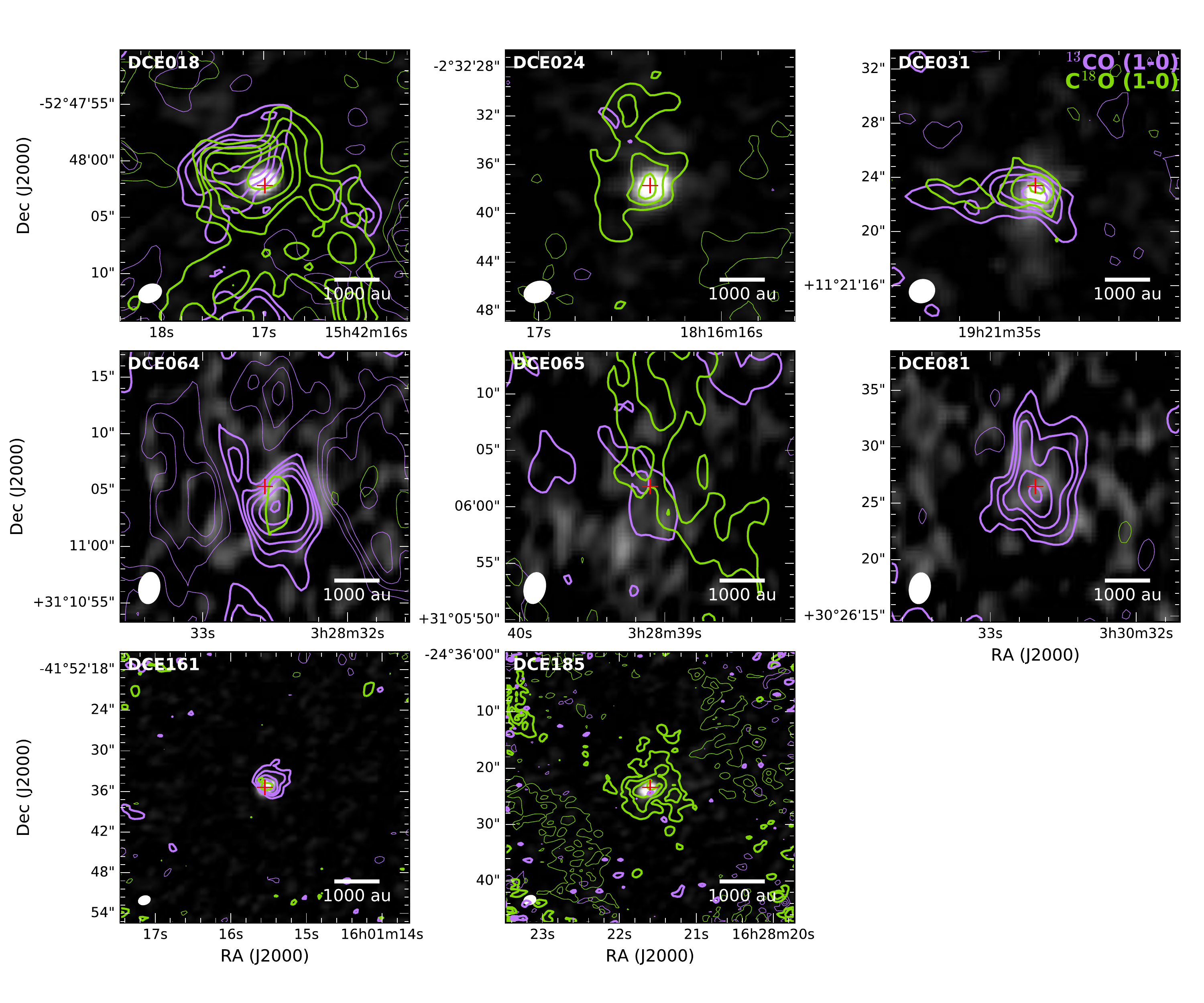}
\caption{Integrated intensity maps of C$^{18}$O (1--0) and $^{13}$CO (1--0) overlaid on the 107.5GHz dust continuum 
emission maps toward the target VeLLOs.
The green and purple contours represent the C$^{18}$O (1--0) and $^{13}$CO (1--0) line emission, respectively, 
with contour levels at 3, 5, 7, 10, and 15$\sigma$.
The thin contours show the negative levels for each line with the corresponding colors.
The red plus signs indicate the infrared source positions from the {\it Spitzer} Space Telescope.}
\label{fig:cont}
\end{figure*}

\section{RESULTS}
\label{sec:results}
\subsection{Continuum emission}
The continuum emission is robustly detected toward the infrared sources in 
DCE018, 024, 031, 161, and 185 with a signal to noise ratio (S/N) $>$ 25, and it is 
marginally detected in the remaining three sources with a S/N between 3 and 5 (Table \ref{tab:gauss}).
The continuum images, at least for the five sources with firm detections, 
show no signs of multiplicity at the angular resolution of the observations 
of several hundred au (Figures \ref{fig:cont} and \ref{fig:DCE018}-\ref{fig:DCE185} ).  This suggests that these VeLLOs are single protostellar systems. 
We fit the continuum maps with a 2-dimensional Gaussian function using the CASA task imfit.
Table \ref{tab:gauss} lists the deconvolved source sizes and respective position angles.
Given the source distances (\citealp[Table \ref{tab:targets};][]{ev09}), the physical sizes 
from the Gaussian fits range from 60 to 500~au.
The major (long) axes are approximately perpendicular to the outflow axes in the four sources 
with outflow detections from the literature 
(\citealp[DCE064:][]{hs17}; \citealp[DCE024:][]{ki11,hs17}; \citealp[DCE185:][]{st06,ba10,ma13,hs16,hs17}; 
\citealp[DCE031:][]{du10a}).
For DCE161, the elongated structure ($0\farcs9\times0\farcs2$, $135\times30$~au) has a 
position angle of 158$\arcdeg$ that is consistent with that (156$\arcdeg$) found for the 
circumstellar disk by \citet{ans16} using resolved 890~$\mu$m continuum emission 
(J16011549-4152351 in the literature).

\begin{figure*}
\centering 
\includegraphics[width=\textwidth]{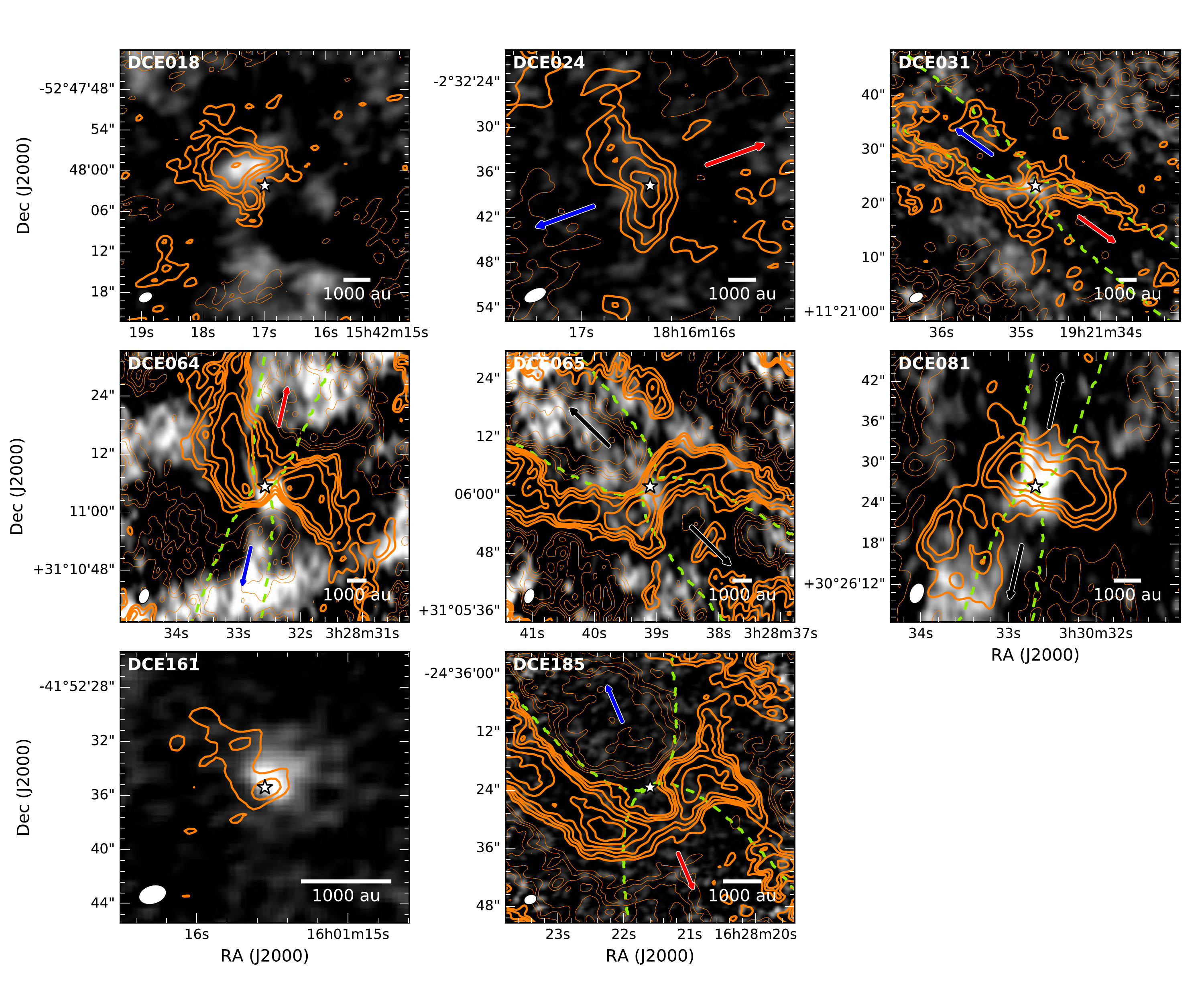}
\caption{Integrated intensity maps of N$_2$H$^+$ (1--0) (orange contours) overlaid on the $^{13}$CO (1--0) integrated intensity maps.
The contour levels are 3, 5, 7, 10, and 15$\sigma$. The thin contours show the negative levels.
The central white stars indicate the infrared source positions from the {\it Spitzer} Space Telescope.
The red and blue arrows represent the outflow directions from the literature, and the hollow arrows show 
the guessed outflow orientations based on the morphology of N$_2$H$^+$ emission whereas the dashed green 
lines show the guessed outflow cavity.}
\label{fig:n2hp_13co}
\end{figure*}

The 3~mm continuum emission in DCE065 peaks at $\approx4\arcsec$ south-east from the 
infrared source and coincides with the N$_2$H$^+$ peak (Figure \ref{fig:DCE065}).
Although the S/N is low ($\approx4$), it is consistent with the SMA 1.3~mm continuum map at a resolution of $4\farcs4\times3\farcs4$
(\citealp{hu10}; see Figure \ref{fig:DCE065} in appendix \ref{sec:tau}), 
which reveals two distinct components.
This suggests that DCE065 may host a binary system with a projected separation
of 5\farcs7 (1400~au).

\subsection{$^{13}$CO (1--0), C$^{18}$O (1--0), and C$^{17}$O (1--0) maps}
\label{sec:co}
Figure \ref{fig:cont} shows the integrated intensity maps of C$^{18}$O (1--0) and $^{13}$CO (1--0) 
overlaid on the continuum images.
In order to avoid contamination from the protostellar outflows, the integrated velocity ranges are 
set to be quite narrow (Table \ref{tab:obs}).
Because C$^{17}$O (1--0) emission is only marginally detected in DCE018 and DCE185 even when integrating 
over all three hyperfine components, we do not plot it in Figure \ref{fig:cont}.
The negative contours in Figure \ref{fig:cont} suggest that these line intensity maps suffer from spatial 
filtering.  
This is especially true for the $^{13}$CO maps (e.g. DCE064).
The missing flux is likely due to emission from structures larger than about 
25$\arcsec$ ($\sim$3000 -- 8000~au) which is the largest angular scale
that can be probed by our data.

\begin{figure*}
\centering
\includegraphics[width=\textwidth]{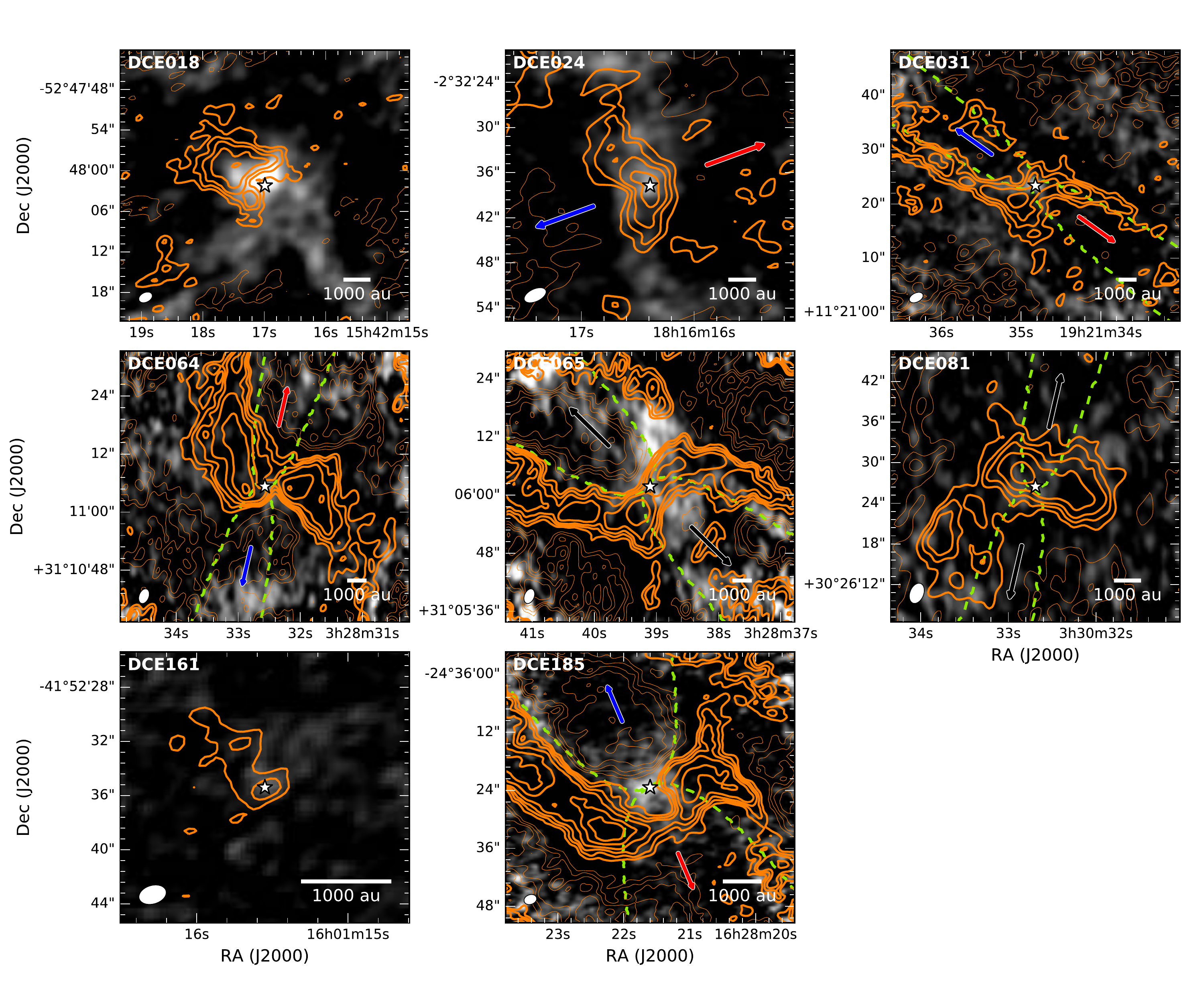}
\caption{Same as Figure \ref{fig:n2hp_13co} but with C$^{18}$O (1--0) maps in gray scale instead of $^{13}$CO (1--0) maps.}
\label{fig:n2hp_c18o}
\end{figure*}

C$^{18}$O and $^{13}$CO are detected (S/N$>$3) toward all sources except for 
$^{13}$CO in DCE024 and C$^{18}$O in DCE081 (Table \ref{tab:gauss}).
The $^{13}$CO emission in DCE185 is only marginally detected (S/N $\sim$ 3.3) 
in a very small area toward the continuum source in Figure \ref{fig:cont}.
For DCE024, we integrate a narrow velocity range covering the C$^{18}$O line emission 
where $^{13}$CO shows a self-absorption dip. 
This feature may originate from optically thick $^{13}$CO emission.
In DCE081, C$^{18}$O emission is undetected (S/N$\sim$2) in the integrated intensity map, 
but is marginally detected in the spatial intensity profiles (see section \ref{sec:profile}).
All the detected CO isotopologue emission peaks at the continuum sources 
(within the beam size) which coincides with the infrared sources.  Emission from $^{13}$CO in DCE018 is an exception to this.
The $^{13}$CO emission in DCE018 peaks 2\farcs4 away from the continuum position 
and is likely to be contaminated by the outflow.
In DCE081, although the $^{13}$CO emission peaks at the continuum/infrared source, 
the extended $^{13}$CO emission is likely to be affected by the outflow cavity in 
a nearly north-south direction (Figure \ref{fig:n2hp_13co}).
More details about the outflows will be discussed in a separate paper.

We fit a 2-dimensional Gaussian to estimate the emitting area of the gas-phase CO isotopologues.
Table \ref{tab:gauss} lists the deconvolved source sizes and the position angles of the 
long axes (east from north) of the C$^{18}$O and $^{13}$CO emission.
For the detected sources, the CO isotopologue emission always occupies a larger area 
than that of the continuum emission.

\subsection{N$_2$H$^+$ (1--0) maps}
\label{sec:n2h}
We integrated emission from all seven hyperfine components of N$_2$H$^+$ over the 
velocity range listed in Table \ref{tab:obs}.
N$_2$H$^+$ is detected toward all eight targets.  Compared with the CO isotopologue emission, N$_2$H$^+$ shows more extended emission 
from the envelopes (Figures \ref{fig:n2hp_13co} and \ref{fig:n2hp_c18o}).
The negative contours in the N$_2$H$^+$ integrated intensity maps suggest 
that there is significant missing flux. 
Taking DCE185 as an example, the ALMA-IRAM 30\,m combined N$_2$H$^+$ integrated intensity map has a higher peak by a factor of $\sim$1.5 and about 87\% of flux is missing within a radius of $\sim$15$\arcsec$. 
Note that, given the complicated structure, 
the estimated missing flux is highly dependent on the selected area.

\tabletypesize{\scriptsize}
\begin{deluxetable*}{ccccccccccc}
\hspace{-10pt}
\tabletypesize{\tiny}
\tablecaption{Parameters of 2-d Gaussian fits}
\tablehead{ 
&\multicolumn{3}{c}{$^{13}$CO (1--0)}
&\multicolumn{3}{c}{C$^{18}$O (1--0)}
& \multicolumn{3}{c}{Continuum}\\
\cmidrule(lr){2-4}
\cmidrule(lr){5-7}
\cmidrule(lr){8-10}	
& \colhead{size/P.A.}
& \colhead{peak}	
& \colhead{S/N}
& \colhead{size/P.A.}
& \colhead{peak}	
& \colhead{S/N}
& \colhead{size/P.A.}	
& \colhead{peak}	
& \colhead{S/N}\\
& (\arcsec)/(degree)
& mJy\,beam$^{-1}$\,km\,s$^{-1}$
&
& (\arcsec)/(degree)
& mJy\,beam$^{-1}$\,km\,s$^{-1}$
&
& (\arcsec)/(degree)
& $\mu$Jy\,beam$^{-1}$
&
}
\startdata 
DCE018	& 7.1$\pm$0.8$\times$4.5$\pm$0.5 (121\arcdeg)& 34.2	& 12.5& 7.9$\pm$0.9$\times$5.7$\pm$0.6 (106\arcdeg)& 51.4	& 19.6& 0.6$\pm$0.2$\times$0.5$\pm$0.2 (176\arcdeg)& 2003	& 43.1 \\
DCE024	& - & 10.4	& 2.9& 4.9$\pm$0.5$\times$3.6$\pm$0.4 (161\arcdeg)& 24.8	& 8.0& 1.4$\pm$0.3$\times$1.3$\pm$0.4 (31\arcdeg)& 1839	& 36.1 \\
DCE031	& 3.7$\pm$0.2$\times$1.7$\pm$0.1 (67\arcdeg)& 47.5	& 12.7& 3.6$\pm$0.5$\times$2.0$\pm$0.4 (72\arcdeg)& 26.2	& 7.4& 2.5$\pm$0.3$\times$1.2$\pm$0.3 (165\arcdeg)& 1439	& 26.7 \\
DCE064	& 4.0$\pm$1.4$\times$3.7$\pm$1.6 (159\arcdeg)& 80.1	& 20.6& - & 20.0	& 4.7& 4.5$\pm$2.0$\times$0.8$\pm$0.6 (168\arcdeg)& 324	& 5.4 \\
DCE065	& - & 29.6	& 4.8& -& 28.3	& 5.1& 9.1$\pm$2.2$\times$7.6$\pm$1.9 (160\arcdeg)& 233	& 4.3\tablenotemark{a} \\
DCE081	& 7.7$\pm$1.1$\times$4.6$\pm$0.7 (160\arcdeg)& 77.8	& 16.4& - & 14.0	& 2.9& 12.1$\pm$3.2$\times$3.8$\pm$1.2 (37\arcdeg)& 212	& 3.4 \\
DCE161	& 3.8$\pm$0.5$\times$3.4$\pm$0.4 (38\arcdeg)& 44.9	& 12.6& - & 13.4	& 4.0& 0.9$\pm$0.1$\times$0.2$\pm$0.2 (158\arcdeg)& 2760	& 65.7 \\
DCE185	& - & 11.7	& 3.3& 11.0$\pm$0.7$\times$8.4$\pm$0.5 (113\arcdeg)& 35.5	& 9.5& 1.2$\pm$0.2$\times$0.8$\pm$0.2 (127\arcdeg)& 2320	& 48.6 
\enddata
\tablecomments{The Gaussian sizes are deconvolved sizes (FWHM) toward the sources detected with a 
signal-to-noise ratio higher than 7. 
The numbers in parentheses represent the position angles of the major axis from north through east.}
\tablenotetext{a}{The peak continuum intensity of DCE065 is about 4\arcsec south-east of the infrared source.}
\label{tab:gauss}
\end{deluxetable*}

All the N$_2$H$^+$ peak positions are offset from the infrared continuum sources, except for DCE161 and DCE024.
Apart from these two objects, N$_2$H$^+$ is likely depleted toward the center of the envelope where 
the CO isotopologue emission peaks.
The observed anticorrelation between N$_2$H$^+$ and CO emission, as seen in \citet{jo04} and \citet{ber02}, 
suggests that N$_2$H$^+$ is destroyed by CO through the well-known gas-phase chemical reaction \citep{ca12}. 
This property makes N$_2$H$^+$ a robust tracer of the CO snow line.
For DCE161 and DCE024, emission from both N$_2$H$^+$ and the CO isotopologues peaks
at the same position as the infrared/continuum source.
In DCE161 (J16011549-4152351) a transition disk has been identified by \citet{va16} based on fitting of the 
spectral energy distribution. 
Subsequently, \citet{ans16} found a disk with a dust mass of 0.061 $M_{\rm Jup}$ and gas mass 
of 6.7 $M_{\rm Jup}$ using resolved ALMA 890 $\mu$m observations.
We speculate that emission from the CO isotopologues and N$_2$H$^+$ traces
the high-density region in the disk of DCE161.
We hereafter remove DCE161 from our study of episodic accretion because it is most likely 
a more evolved source with a dissipating envelope.  This prevents us from measuring the radius of the CO snow line using N$_2$H$^+$.
In the case of DCE024, there are two possibilities to explain the common peak position for N$_2$H$^+$ and C$^{18}$O:
(1) the emission from N$_2$H$^+$ and C$^{18}$O is in fact spatially separated but projected onto the same region along the line of sight, and
(2) the spatial resolution of our data is too low to resolve the anti-correlation between emission from N$_2$H$^+$ and the CO isotopologues.

The N$_2$H$^+$ integrated intensity maps reveal flattened envelopes in six out of eight targets 
(except for DCE018 and DCE161), four of which have major axes perpendicular to their outflow axes 
(Figure \ref{fig:n2hp_13co}) as determined from the literature 
(\citealp[DCE024, 064, and 185 in][]{hs17} and \citealp[DCE031 in][]{du10a}).
 Although the presence of outflows are not yet reported for DCE065 and DCE081, 
the flattened envelopes are likely perpendicular to potential outflows. 
In DCE081, the $^{13}$CO emission is likely associated with an outflow 
(see section \ref{sec:co}) orientated perpendicular to the major axis of the envelope traced in N$_2$H$^+$ emission.
In DCE065, the N$_2$H$^+$ emission shows two ``U-shaped shells'' (or ``H-shaped''). This emission morphology, as seen in the sources with a detected outflow (DCE031, 064, 065, 185), 
likely originates from an outflow-compressed envelope (Figures \ref{fig:n2hp_13co} and \ref{fig:n2hp_c18o}).
As a result, N$_2$H$^+$ emission would appear to highlight the 
outer shells, while that from the CO isotopologues traces the outflow entrained gas or the outflow cavity walls.
Therefore, if we adopt the shells as the outflow orientations in DCE081 and DCE065, 
the flattened envelopes, including the four sources with detected outflows, 
are roughly perpendicular to the outflow axes in all six flattened envelopes as traced in emission from N$_2$H$^+$.

\begin{figure*}
\centering 
\includegraphics[width=\textwidth]{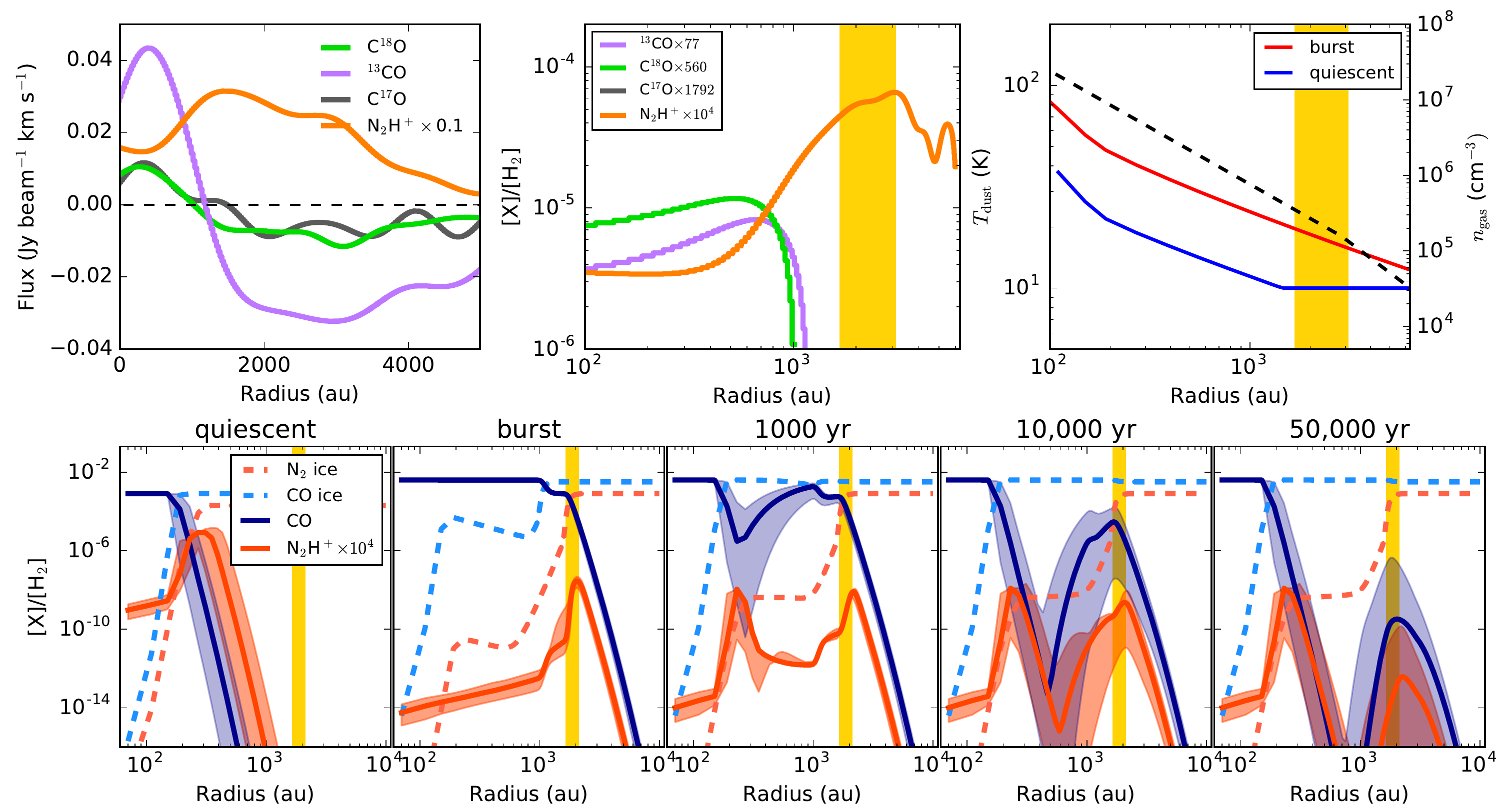}
\caption{
DCE064 is shown here as an example of the analysis used in this paper. 
\textit{Top left:} Intensity profiles of the observed molecules taken along the cut shown in 
Figures \ref{fig:DCE018} -- \ref{fig:DCE185}. 
\textit{Top center:} Abundance profiles of the observed molecules. 
\textit{Top right:} Model of temperature and density profiles for the source; 
the red and blue lines show the temperature profiles corresponding to the burst and current luminosities, respectively. 
\textit{Bottom panels:} Time-dependent chemical models of CO and N$_2$H$^+$ abundance profiles for the quiescent, 
burst and post-burst stages. 
The yellow vertical area represents the observed N$_2$H$^+$ depletion radius, which suggests that DCE064 likely experienced 
a past burst within $\sim$10,000~yr.}
\label{fig:demo}
\end{figure*}

\section{ANALYSIS}
Figure \ref{fig:demo} summarizes the analysis we perform in this section using the results for DCE064 as an example. 
The details of each step are described in the following subsections. 
We first take the intensity profiles of the molecular line emission along a cut perpendicular 
to the outflow axis of each VeLLO. 
Then, abundance profiles for N$_2$H$^+$ and CO are calculated assuming a density profile in the envelope. 
Third, the radius of the CO snow line is determined by modeling the N$_2$H$^+$ abundance peak position. 
Using the radius of the CO snow line, the luminosity of the central star during the past accretion burst was 
derived assuming a CO sublimation temperature of 20~K.
Finally, the modeled temperature profiles are coupled with a time-dependent full chemical network to model 
the evolution of the abundance of CO and N$_2$H$^+$ through the envelope of each source in time.

\label{sec:analysis}
\subsection{Comparison of intensity profiles}
\label{sec:profile}
\begin{figure*}
\centering 
\includegraphics[width=\textwidth]{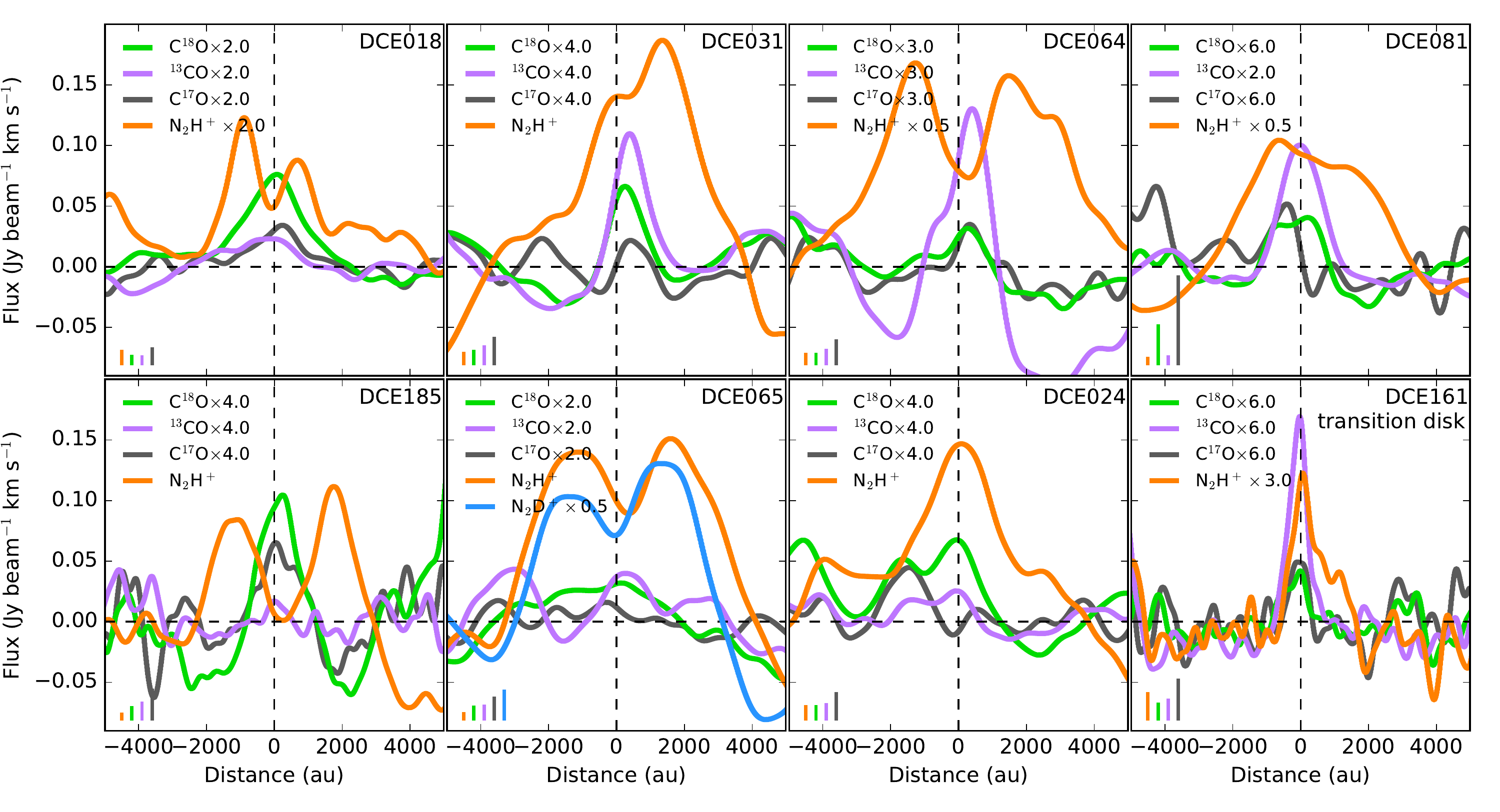}
\caption{Intensity profiles of N$_2$H$^+$ (1--0) (orange), C$^{18}$O (1--0) (green), $^{13}$CO (1--0) (purple), 
and C$^{17}$O (1--0) (gray) along the cuts shown in Figures \ref{fig:DCE018}-\ref{fig:DCE185}. 
The vertical dashed lines indicate the infrared source positions \citep{du08}. 
The multiplicative factors used to scale the profiles are shown in the upper left corner.
The color bars in the lower left corner indicate the 1$\sigma$ rms noise level of the profiles in the same color. 
The sources are sorted by the three categories in section \ref{sec:profile}.
The N$_2$D$^+$ (3--2) profile in DCE065 corresponds to SMA observations of \citet{hu10}.}
\label{fig:profile}
\end{figure*}

To compare the spatial distribution of emission from N$_2$H$^+$ and CO, 
we plot the intensity profiles of N$_2$H$^+$ and the CO isotopologues (Figure \ref{fig:profile}).
These profiles are obtained using the integrated intensity maps shown in Figures \ref{fig:cont} 
to \ref{fig:n2hp_c18o} along the cuts shown in Figures \ref{fig:DCE018} to \ref{fig:DCE185}.
The cuts cross the source centers and have a width of 4\arcsec, about $1-3$ times the beam size, 
which should only marginally affect the profiles.
The position angles of the cuts are selected based on the following criteria.
For the four sources with outflow detections in the literature, the cuts are taken perpendicular 
to the outflow axes (DCE024, 064, and 185 from \citealt{hs17}; and DCE031 from \citealt{du10a}). 
For the two sources with potential outflows (DCE065 and DCE081), we take the cuts perpendicular 
to the assumed outflow axes that are inferred from our N$_2$H$^+$ and CO maps 
(see section \ref{sec:n2h}; Figure \ref{fig:n2hp_13co}).
DCE018 and DCE161 show no clear indication of the existence of protostellar outflows nor their orientations in the plane of the sky.
In DCE018, both $^{13}$CO and C$^{18}$O maps show elongated structures with the long axis 
roughly aligned from the north-west to the south-east (Table \ref{tab:gauss}).
The N$_2$H$^+$ map shows an arc-like structure directed toward the south-west which surrounds 
the continuum emission (Figure \ref{fig:DCE018}).
We choose the cut across the two arms of the N$_2$H$^+$ arc-like structure from the north-west 
to the south-east in order to feature the two N$_2$H$^+$ peaks in the intensity profile (Figure \ref{fig:profile}).
In DCE161, we choose a cut along the major axis of the continuum emission (Gaussian fit in Table \ref{tab:gauss}). 
The position angle of the cut is consistent with the long axis of the protostellar disk found by \citet{ans16} 
from resolved continuum observations at 890~$\mu$m.

Figure \ref{fig:profile} shows the spatial intensity profiles of N$_2$H$^+$ (1--0), 
$^{13}$CO (1--0), and C$^{18}$O (1--0) obtained along the cuts described above.
Based on these profiles, we categorize the targets into three types 
(excluding DCE161, see section \ref{sec:n2h}):
\begin{enumerate}[(a)]
\item Detection of N$_2$H$^+$ depletion toward the center where CO evaporates: 
DCE018, 031, 064, 081 and 185 enter into this category.
N$_2$H$^+$ is destroyed by gaseous CO and therefore highlights the CO snow line.
\item Detection of N$_2$H$^+$ depletion toward the center with very weak CO emission: 
in DCE065, N$_2$H$^+$ depletion is clearly seen, 
but the C$^{18}$O (1--0) and $^{13}$CO (1--0) lines are very weak toward the center as observed with ALMA, 
and both C$^{18}$O (2--1) and $^{13}$CO (2--1) lines are undetected with SMA \citep{hu10}.
A possible explanation is that DCE065 has experienced a burst but the inherently low abundance
CO gas is consumed by reaction with N$_2$H$^+$.
An alternative possibility is the freeze-out of N$_2$, the parent molecule of N$_2$H$^+$, in the high density central region 
as interpreted in IRAM 04191 by \citet{be04}.
Furthermore, the SMA observations of \citet{hu10} reveal N$_2$D$^+$ depletion toward the center 
with a radius similar to that of N$_2$H$^+$ (Figures \ref{fig:profile}).
This result is different from the case of L1157, in which \citet{to13} find a depletion radius 
of N$_2$D$^+$ larger than that of N$_2$H$^+$. 
This difference in radius is explained by invoking the fact that, after an accretion burst, 
N$_2$D$^+$ takes a longer time to recover than N$_2$H$^+$ due to the lack of abundant 
H$_2$D$^+$ when T $\gtrsim$ 20 K \citep{ca12}.
As a result, either DCE065 has experienced a very recent burst, and thus the radius difference 
has not yet become significant, or both N$_2$H$^+$ and N$_2$D$^+$ are depleted due to the freeze-out 
of N$_2$ in the high density central region.
\item Extended N$_2$H$^+$ and C$^{18}$O emission with a similar peak position: 
in DCE024, the N$_2$H$^+$ and C$^{18}$O emission have a common peak position toward the 
infrared sources. 
One possibility is that the overlap in peak position between N$_2$H$^+$ and C$^{18}$O is 
due to a projection effect.
Alternatively, an anti-correlation between C$^{18}$O and N$_2$H$^+$ may exist on spatial
scales smaller than those probed by our observations.
\end{enumerate}

\subsection{Molecular Column Densities}
\label{sec:col}
\begin{figure*}
\centering
\includegraphics[width=\textwidth]{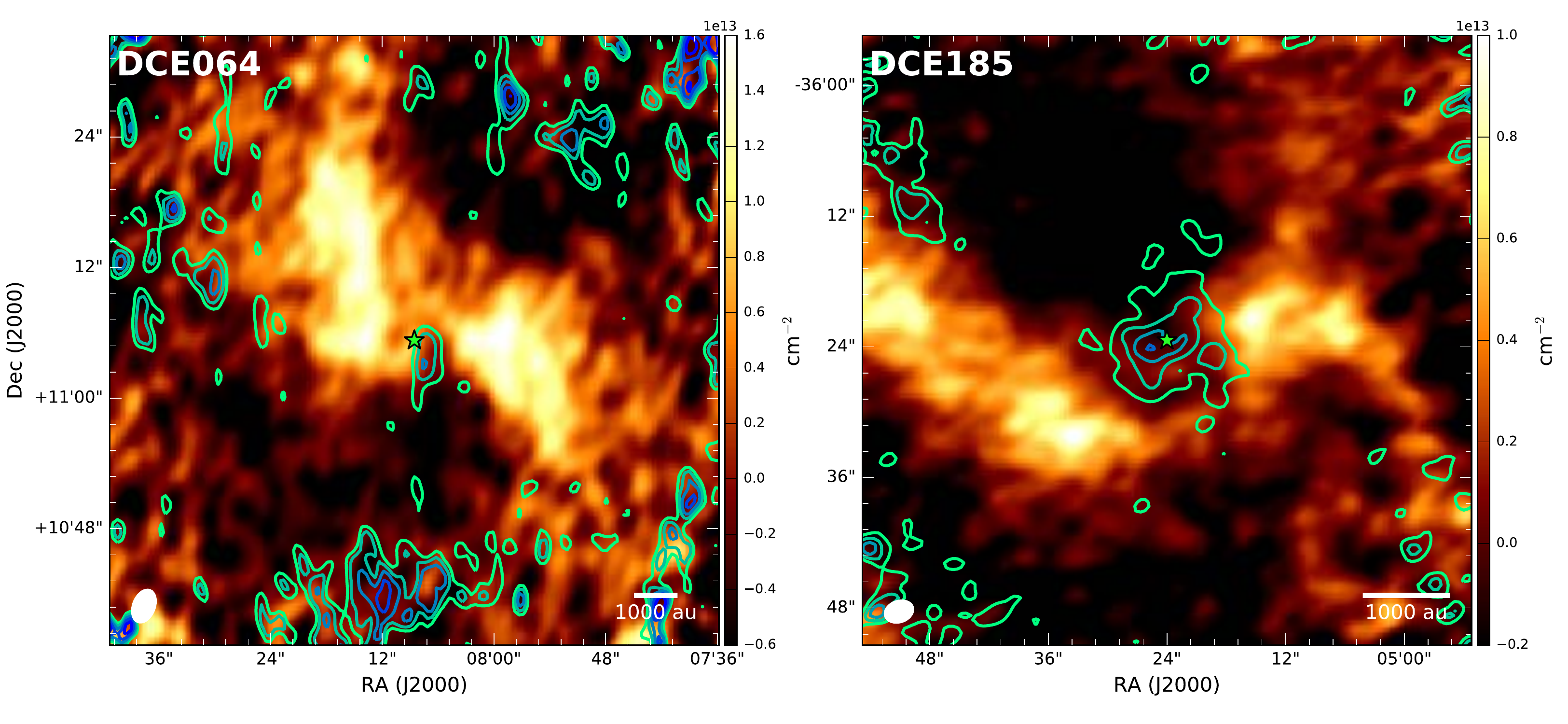}
\caption{CO column density map (contours) overlaid on that of N$_2$H$^+$ (color scale) in DCE064 (left) and DCE185 (right). 
The CO column density is obtained by scaling the C$^{18}$O column density map with the isotopic ratio, 
560 \citep{wi94}. 
The contour levels are 1, 1.5, and 2.0 $\times 10^{17}$~cm$^{-2}$ for DCE064 and 3, 5, 7, and 9 $\times 10^{17}$ cm$^{-2}$ for DCE185.}
\label{fig:column_maps}
\end{figure*}

We derive the CO and N$_2$H$^+$ column density maps toward all eight VeLLOs. 
Figure \ref{fig:column_maps} shows the column density maps derived for 
DCE064 and DCE185 as examples. 
The N$_2$H$^+$ column density maps of the eight VeLLOs are further shown in 
Figures \ref{fig:DCE018} to \ref{fig:DCE185}. 
The details of the calculations are described below for CO (section \ref{sec:col_co}) 
and N$_2$H$^+$ (section \ref{sec:col_n2hp}).

\subsubsection{Column Density of CO - Assumption of optically thin CO isotopologues emission}
\label{sec:col_co}
\begin{figure*}
\includegraphics[width=0.95\textwidth]{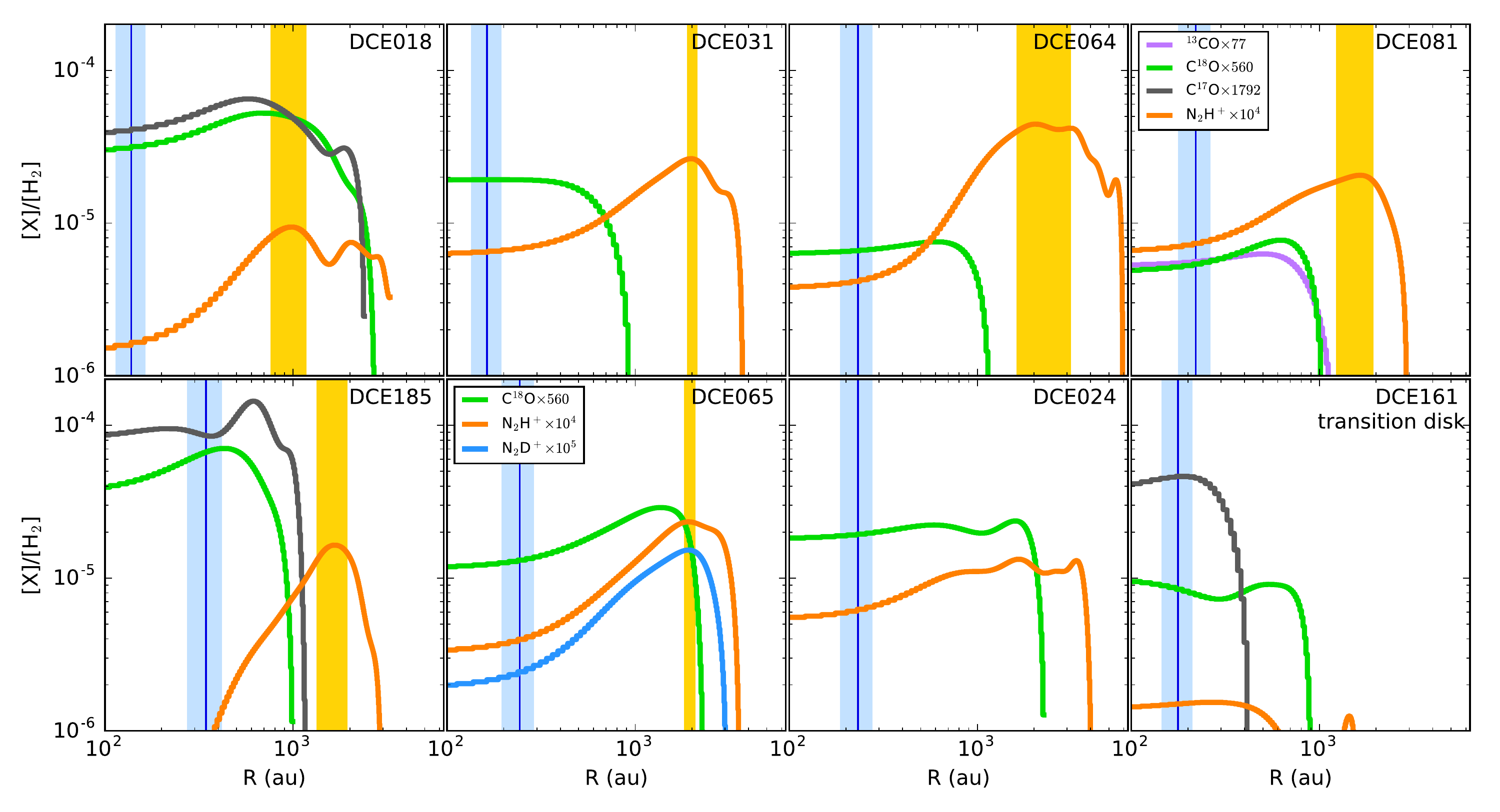}
\caption{Abundance profiles of N$_2$H$^+$ (orange) and CO along the same cuts as in Figure \ref{fig:profile}. 
The abundances represent the average values of both sides.
The CO abundance is calculated by multiplying the isotopic ratio with the abundances of C$^{18}$O (green), 
$^{13}$CO (purple), or C$^{17}$O (gray) estimated under the assumption of optically thin emission.
The yellow area represents the radius of N$_2$H$^+$ abundance peak with its uncertainty 
(width, see appendix \ref{sec:power-law_index}).
DCE024 and DCE161 show no evidence for N$_2$H$^+$ depletion (see section \ref{sec:abundance}).
The blue lines show the location where the dust temperature is predicted to be 20~K based on the current 
bolometric luminosity under the assumption of an envelope mass of 1~$M_\odot$ and the blue area shows the 
range predicted for envelope masses between 0.5 and 3.0~$M_\odot$.}
\label{fig:abundance}
\end{figure*}

We estimate the column density of CO under the assumptions of optically thin emission of C$^{18}$O and 
local thermodynamic equilibrium (LTE) with an excitation temperature of 10~K in our eight targets.
Here we discuss the validity of the assumption of optically thin emission based on the 
intensity ratios between the three CO isotopologues, $^{13}$CO, C$^{18}$O, and C$^{17}$O.
Taking the isotopic ratio, $^{13}$CO/C$^{18}$O $\sim7.3$, in the local ISM \citep{wi94}, 
the C$^{18}$O emission can be considered as optically thin ($\tau<1$) if the intensity ratio 
of $^{13}$CO to C$^{18}$O is larger than $\sim$1.6 (\citealp[see, e.g., eq.\ 7 in][]{sh14}).
Thus, the intensity profiles in Figure \ref{fig:profile} imply that the C$^{18}$O emission 
in DCE031, 064, 081, and 161 is most likely optically thin.
Similarly, given the isotopic ratio of C$^{18}$O/C$^{17}$O $\sim$ 3.2, the intensity ratio of 
C$^{18}$O to C$^{17}$O is larger than $\sim$2.4 when the C$^{18}$O emission is optically thin.
Therefore, the C$^{18}$O emission from DCE024 is likely to be optically thin given the 
non-detection of C$^{17}$O.
The C$^{18}$O emission from DCE065 is assumed to be optically thin, because both $^{13}$CO and 
C$^{18}$O lines are only marginally detected.
For the two sources in which C$^{17}$O is detected, DCE018 and 185, we assume that the emission from both C$^{17}$O and C$^{18}$O is optically thin. We find that the resulting column density ratios of 
C$^{17}$O to C$^{18}$O are broadly consistent with the local ISM abundance ratio 
(see Figure \ref{fig:abundance}).

It is noteworthy that three sources (DCE018, 024, and 185) have C$^{18}$O intensities
larger than that of $^{13}$CO. 
This mostly comes from the optically thick emission and self-absorption of $^{13}$CO which 
suffers from substantial spatial filtering in the interferometric observations.

\subsubsection{Column Density of N$_2$H$^+$}
\label{sec:col_n2hp}
We calculate the N$_2$H$^+$ column density map toward each source using the isolated 
component ($JF_1F$: 101--012) under the assumption of optically thin emission.
We adopt this assumption because some uncertainties prevent us from estimating the optical depth 
accurately from the hyperfine structure.
First, although our data reveal clear N$_2$H$^+$ detections based on the integrated intensity 
(Figure \ref{fig:n2hp_13co}), 
the signal-to-noise ratios are sometimes insufficient to fit the hyperfine spectra 
(Figures \ref{fig:DCE018}-\ref{fig:DCE185}).
Second, our data sets have missing flux problems as shown by the negative contours
in Figure \ref{fig:n2hp_13co}. 
Spatial filtering not only results in underestimating the overall intensity but also affects
the relative intensities of the hyperfine components. 
Therefore, the optical depth ($\tau$) derived from these ratios is also affected, because components 
with different opacities may probe different structures.
Third, hyperfine anomalies due to non-thermal effects \citep{da06,da07,ke10,lo12} could also 
change the ratios between the hyperfine components. Although the optical depth cannot be determined precisely across the whole map, 
we find that the optically thin assumption is reasonable for the isolated 101--012 component in our 
targets (see appendix \ref{sec:tau}). 
We therefore use this component and assumption to derive the column density maps for N$_2$H$^+$.

We calculate the N$_2$H$^+$ column density using the isolated 101--012 component 
(except for DCE161, see the following) under the assumptions of optically thin emission and 
LTE with an excitation temperature of 10~K (see, e.g., Eq.\ 79 in \citealt{ma15}).
The assumed excitation temperature is comparable to the estimated gas temperature from the non-LTE 
analysis with RADEX \citep{va07} of the N$_2$H$^+$ (3--2)/(1--0) ratio spectra in our single-dish 
survey of VeLLOs \citep{hs15} which includes five of our targets. 
The column density will change by factors of 0.3 and 1.7 if we assume excitation temperatures of 
5 and 15~K, respectively.
Figures \ref{fig:DCE018} to \ref{fig:DCE185} show the resulting column density maps.
Two caveats should be kept in mind. Firstly, spatial filtering is significant as indicated by the negative contours in Figure \ref{fig:n2hp_13co}.
Given the largest angular scales ($\sim$25\arcsec, $\sim$3000--8000 au) of the data, the missing flux is 
mostly related to the cloud core and cloud structures ($\gtrsim$20,000 au), which may not significantly 
affect our analysis at the scale of the envelope ($\sim$1000--10,000 au). Secondly, due to the hyperfine anomalies described above, the isolated 101--012 component is usually considerably brighter 
than that expected from the statistical weight (Figures \ref{fig:DCE018} to \ref{fig:DCE185}).
However, in our study of the N$_2$H$^+$ distribution, the absolute column density will not 
significantly affect our conclusions.

For DCE161, the N$_2$H$^+$ emission is too weak if we only integrate the isolated 101--012 component 
(3/27 of the strength of the whole multiplet). 
We thus calculate the column density using the integrated intensity over all hyperfine components. 
We suggest that the assumption of optically thin emission for all seven hyperfine components is reasonable 
because the weak emission ($\lesssim$2~K toward the strongest component, Figure \ref{fig:DCE161}) is unlikely to trace a region with a high column density.

\subsection{Abundance Profiles of N$_2$H$^+$ and CO}
\label{sec:abundance}
In order to calculate the N$_2$H$^+$ abundances relative to H$_2$, we construct a model core with a 
spherical symmetry and a broken power-law density profile. 
The power-law index, $p$, was assumed to be $-1.5$ for the inner free-fall region and $-2.0$ for the 
outer static region \citep{yo05}.
The transition radius from static to free-fall is adopted to be 3000~au based on the dynamical model of 
the well-studied VeLLO, IRAM 04191 \citep{be02}. 
We assume three core masses of 0.5, 1.0, and 3.0 $M_\odot$ within a radius of 10,000~au which are 
comparable to that (0.3--1.0 M$_\odot$) from the single-dish observations toward the four VeLLOs 
in Perseus and Ophiuchus \citep{en08}.
We integrate the density of the material along the line of sight to obtain the H$_2$ column density.
To derive the N$_2$H$^+$ and CO abundances, we divide the corresponding molecular column density 
(section \ref{sec:col}) by the H$_2$ column density.
We then calculate the abundance profiles as a function of radius (Figure \ref{fig:abundance}) 
by averaging over that of the two sides of the protostar along the selected axis.

\tabletypesize{\scriptsize}
\tabcolsep=0.3cm
\begin{deluxetable*}{cccccccccc}
\tabletypesize{\tiny}
\tablecaption{Current and predicted luminosity and the corresponding CO snow line positions and mass accretion rate}
\tablehead{
& $L_{\rm bol}$
& $L_{\rm int}$
& $R_{L_{\rm bol}}$
& $R_{L_{\rm int}}$
& $R_{L_{\rm bol,convolved}}$
& $R_{\rm N_2H^+,peak}$
& $L_{\rm burst}$
& $R_{\rm burst}$
& $\dot{M}_{\rm acc}$
\\
& $L_\odot$
& $L_\odot$
& \colhead{(au)}
& \colhead{(au)}
& \colhead{(au)}
& \colhead{(au)}
& \colhead{($L_\odot$)}
& \colhead{(au)}
& $M_\odot$yr$^{-1}$
}
\startdata
DCE018	& 0.06$\pm$0.01	& 0.04	& 114-164	& 102-145	& 275-311	& 775.0-1149.0	& 1.0-4.0	& 633-836	& (0.6$\pm$0.4)$\times10^{-5}$	\\
DCE024	& 0.20$\pm$0.04	& 0.07	& 186-277	& 122-175	& 422-486	& -	& -	& -	& -	\\
DCE031	& 0.09$\pm$0.03	& 0.04	& 134-194	& 102-145	& 362-410	& 1919.0-2100.0	& 5.1-13.4	& 1527-1593	& (2.3$\pm$1.0)$\times10^{-5}$	\\
DCE064	& 0.20$\pm$0.05	& 0.03	& 186-277	& 94-132	& 428-493	& 1649.0-3075.0	& 3.9-29.1	& 1316-2404	& (4.1$\pm$3.2)$\times10^{-5}$	\\
DCE065	& 0.22$\pm$0.06	& 0.02	& 195-290	& 87-117	& 437-505	& 1850.0-2049.0\tablenotemark{a}	& 4.7-12.7	& 1459-1547	& (2.2$\pm$1.0)$\times10^{-5}$	\\
DCE081	& 0.18$\pm$0.04	& 0.06	& 177-264	& 115-165	& 417-480	& 1249.0-1900.0	& 2.2-10.9	& 961-1426	& (1.6$\pm$1.1)$\times10^{-5}$	\\
DCE161	& $\geqslant$0.11	& 0.08	& 145-212	& 128-185	& 251-290	& -	& -	& -	& -	\\
DCE185	& 0.45$\pm$0.08	& 0.09	& 272-418	& 134-195	& 356-457	& 1362.0-1900.0	& 2.7-10.9	& 1073-1426	& (1.7$\pm$1.0)$\times10^{-5}$
\enddata
\tablecomments{
Col. (1)-(3): Source properties from \citet{du08}.
Col. (4): CO sublimation radius (20K) corresponding to the current luminosity in Col. (2). 
The uncertainty is given by different assumption of envelope mass from 0.5-3.0 $M_\odot$.
Col. (5): CO sublimation radius (20K) corresponding to the internal luminosity in Col. (3).
Col. (6): CO sublimation radius (Col. 4) convolved with the observational beam size.
Col. (7): Measured radius of N$_2$H$^+$ abundance peak, see section \ref{sec:abundance}.
Col. (8): Source luminosity at the burst phase from models, see section \ref{sec:chemical_model}.
Col. (9): CO sublimation radius (20K) corresponding to the outburst luminosity ($L_{\rm burst}$) in Col. (8).
Col. (10): Mass accretion rates estimated based on the outburst luminosity ($L_{\rm burst}$) in Col. (8).}
\tablenotetext{a}{The N$_2$H$^+$ depletion in DCE065 could alternatively come from the freezeout of the parent molecule, N$_2$. 
In such a case, DCE065 did not experience a recent accretion burst.}
\label{tab:result}
\end{deluxetable*}

We obtain the peak radii of the N$_2$H$^+$ abundance ($R_{\rm N_2H^+,peak}$, Table \ref{tab:result}) 
which are used for the modeling in section \ref{sec:chemical_model}.
The derived peak radii may be affected by the assumed H$_2$ density profile because the abundance 
profile depends on the H$_2$ column density. 
In order to test this, we compare the derived $R_{\rm N_2H^+,peak}$ with that calculated by assuming 
pure power-law density profiles with indices of $p=-1.5$ and $p=-2.0$. 
The abundance profiles for the model with $p=-1.5$ are almost the same as those
for the broken power-law, whereas those with $p=-2.0$ have steeper slopes toward the centers 
(Figures \ref{fig:abundance_a}). 
For the model with $p=-2.0$, $R_{\rm N_2H^+,peak}$ is larger by 3--13\% than that with $p=-1.5$. 
This difference is taken as the uncertainty in this measurement (appendix \ref{sec:power-law_index}).
In the case of DCE024, the N$_2$H$^+$ abundance profile drops toward the source center, especially for 
the model with $p=-2.0$. However, the decrease (by a factor of $\sim 2-5$) is likely to be produced by the 
assumed density profile rather than destruction by CO because of the centrally-peaked N$_2$H$^+$ 
intensity profile (Figure \ref{fig:profile}).
Therefore, we measure $R_{\rm N_2H^+,peak}$ only for the sources with N$_2$H$^+$ depletion in the 
intensity profiles (i.e., categories (a) and (b) in section \ref{sec:profile}).

\subsection{Models of temperature profiles}
\label{sec:TD}
Given the bolometric luminosities (Table \ref{tab:targets}),
we use the one-dimensional radiative transfer code DUSTY \citep{iv99} to model the temperature profiles of our VeLLOs.
We use the modeled cloud density distribution described in section \ref{sec:abundance} which is scaled 
to the envelope masses of 0.5, 1.0, and 3.0~$M_\odot$ within 10,000~au.
Assuming a gas-to-dust mass ratio of 100 and a dust opacity of 1063.8~cm$^2$~g$^{-1}$ at 8~$\mu$m \citep{os94}, 
we calculate the dust temperature profile with the radiative transfer modeling.
As a result, we find that the CO snow line, assumed to be at a temperature of 20~K, 
is located at a very small radius of $R_{L_{\rm bol}}\sim 150-450$~au when
considering only the bolometric luminosity at the present time (Table \ref{tab:result}).
These values imply that the faint central source of a VeLLO can evaporate CO and in turn 
destroy N$_2$H$^+$ in a region that is not resolved by our observations.
Table \ref{tab:result} lists the CO sublimation radii predicted by the radiative transfer model 
in two cases: one in which the luminosity assumed to be equal to the internal luminosity and another in which it is assumed to be equal to the bolometric luminosity.
Furthermore, for a better comparison to the observations, the model temperature profiles are
convolved with a Gaussian with the same size as the observed beam ($R_{L_{\rm bol, convolved}}$ 
in Table \ref{tab:result}).

\subsection{Chemical modeling}
\label{sec:chemical_model}
We model the observed N$_2$H$^+$ abundance profiles at the post-burst phase for the six sources 
with N$_2$H$^+$ depletion and at the quiescent phase for all eight sources.
We adjust the luminosity used as an input for the DUSTY simulation so that the temperature profile 
leads to a N$_2$H$^+$ peak located at a radius $R_{\rm N_2H^+,peak}$ as determined from the observations 
(section \ref{sec:abundance}). 
The best-fit outburst luminosity ($L_{\rm burst}$) and its corresponding CO sublimation radius 
($R_{\rm burst}$ at 20 K) are listed in Table \ref{tab:result}.
A lower limit of 10~K is set for the temperature profiles to mimic the external heating from the 
interstellar radiation field for all models except for DCE024. 
To better reproduce the extended N$_2$H$^+$ emission toward DCE024, 
we take 15~K as a lower limit. 
This temperature is slightly lower than the N$_2$ sublimation temperature 
of 18~K at a density of 10$^5$~cm$^{-3}$ given a binding energy of 955~K (see below).

The N$_{2}$H$^{+}$ abundance profiles are modeled using a full time-dependent deuterated chemical network. 
The network is based on the UMIST database for Astrochemistry RATE06 version \citep{wo07} extended 
with deuterium fractionation reactions \citep{mc13}. 
The network includes gas phase chemistry, gas-grain reactions (freeze-out, thermal desorption, 
and cosmic-ray-induced photodesorption), and grain-surface reaction \citep{ha92}.
The rate coefficient for freeze-out (gas to ice) is obtained by
\begin{equation}
k_{\rm acc} = \sigma_{\rm d}~\langle v_{i}\rangle~n_{\rm d}~~(s^{-1})
\end{equation}
while the rate coefficient for thermal desorption (ice to gas) is given by
\begin{equation}
k_{\rm desp} = \nu_{0}~{\rm exp}\left(\frac{-E_{\rm desp}(i)}{k~T_{\rm d}}\right) ~~(s^{-1})
\end{equation}
where $\sigma_{\rm d}$ is the dust grain cross section with an assumed grain size of 0.1~$\mu$m, 
$\langle v_{i}\rangle$ is the thermal velocity of species $i$, 
$n_{\rm d}$ is the dust number density, 
$E_{\rm desp}(i)$ is the molecular binding energy, and 
$\nu_{0}$ is the characteristic vibration frequency of the absorbed species. 
The characteristic vibration frequency is given by:
\begin{equation}
\nu_{0} = \sqrt{\frac{2~N_s~E_{desp}(i)}{\pi^2~m_i}}
\end{equation}
where $N_s = 1.5\times10^{15}$~cm$^{-2}$ is the number density of surface sites on the dust grains
and $m_i$ is the mass of the species $i$.
Table \ref{tab:chemistry} lists the relevant reactions used in the network, 
along with the parameters needed to calculate the rate coefficient for two-body reactions and cosmic ray ionization. 
For a two-body reaction, the rate coefficient is given by
\begin{equation}
\label{eqratecoeff2b}
k = \alpha \left(\frac{T}{300}\right)^\beta {\rm exp}\left(-\frac{\gamma}{T}\right) ~\rm cm^{3}~s^{-1}
\end{equation}
with the temperature denoted by $T$. For cosmic ray ionization, the rate coefficient is described by
\begin{equation}
\label{eqratecoeff1b}
k = \zeta ~\rm s^{-1}
\end{equation}
Table \ref{tab:chemistry} lists the parameters $\alpha$, $\beta$, $\gamma$ and $\zeta$ needed to calculate the rate coefficient of each reaction.

The chemical network is coupled with a source density and temperature profile as a function of radius.
The abundances of CO and N$_{2}$ relative to H$_2$, $X_{\rm CO}$ and $X_{\rm N_2}$, are fixed at a 
constant value of 4 $\times$ 10$^{-4}$ and 1 $\times$ 10$^{-4}$, respectively.
The binding energy of CO is assumed to be 1150~K (corresponding to a sublimation temperature of $\sim$20 K), 
and the N$_2$ binding energy is taken as 0.83 times that of CO (955 K) \citep{an16}. 
This binding energy is similar to that derived experimentally \citep{bi06,fa16}. 
This model set up calculates the sublimation radius of CO and N$_2$ 
for a given temperature profile.  The aim is to reproduce the N$_2$H$^+$ ring with the inner radius set by the destruction of N$_2$H$^+$ by gaseous CO and the outer radius determined 
by the freeze-out of N$_2$.

\begin{figure*}
\includegraphics[width=0.9\textwidth]{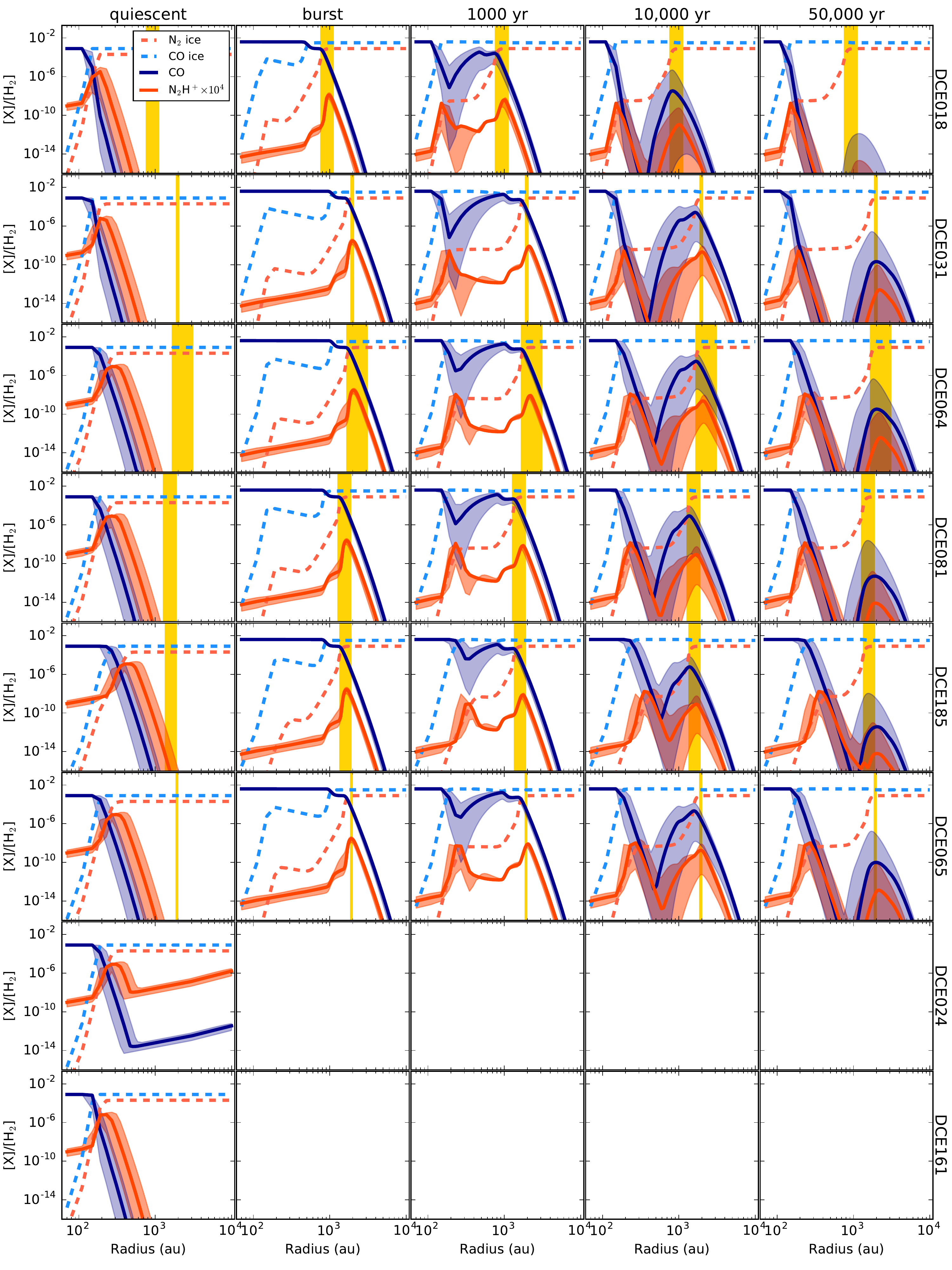}
\caption{Time-dependent models of the abundance profiles.
First column: profiles during the quiescent phase with $L=L_{\rm bol}$. 
Second column: profiles during the burst phase with $L=L_{\rm burst}$ (Table \ref{tab:result}) assumed to last for 100 yr. 
Third to fifth columns: profiles 1,000, 10,000, and 50,000~yr after the burst phase, with $L=L_{\rm bol}$. 
The yellow area is the same as in Figure \ref{fig:abundance}.
The blue and orange solid lines show the gaseous CO and N$_2$H$^+$ abundance profiles, while the blue and orange 
dashed lines indicate that of CO and N$_2$ sticking onto dust grains. 
The colored areas represent the range of profiles resulting from assuming envelope masses between 
0.5~$M_\odot$ and 3.0~$M_\odot$. 
Synthetic abundance profiles smoothed to the angular resolution of the observations are shown in Figure~\ref{fig:model_con}.}
\label{fig:model}
\end{figure*}

To reproduce the chemical evolution of episodic accretion, the quiescent, burst and post-burst phases 
are modeled for each object. 
The results of this time-dependent chemical model are shown in Figure \ref{fig:model}.
For the quiescent phase, the source density and temperature profile with the current measured luminosity are used.
CO and N$_{2}$ are initially set to be completely frozen onto the dust grains 
($X_{\rm gCO}$ = 4 $\times$ 10$^{-4}$, $X_{\rm gN_2}$ = 1 $\times$ 10$^{-4}$), 
and the model is evolved for 10$^{6}$ yr.
This phase is followed by the accretion burst, where the source profile is switched to that with 
an increased luminosity (i.e. $L_{\rm burst}$ in Table \ref{tab:result}).
The model is then evolved for 100 and 200~yr \citep{vo05}, which is the expected duration of an 
accretion burst. 
However, the duration of the burst does not appear to have a significant effect on the 
distribution or abundance of N$_2$H$^+$.
Finally, the source profile is again switched back to the luminosity of the quiescent phase 
in order to model the evolution of the gas post accretion burst.
The post-burst model is evolved up to 10$^{5}$~yr, and the results at 1000~yr, 10,000~yr, 
and 50,000~yr after the burst are compared in order to see how the peak abundance positions of CO and 
N$_2$H$^+$ shift with time.
Shortly after the accretion burst has ended, CO starts to drop at a radius of about 400~au, 
causing N$_2$H$^+$ to generate a second inner peak at the radius of the CO drop.
With time, the second inner peak of N$_2$H$^+$ becomes more prominent, while the gas in the 
outer peak starts to freeze-out onto the dust grains again.
Thus, starting from the inner high-density region, N$_2$H$^+$ progressively resumes its initial position during the quiescent, pre-burst phase.

\begin{figure*}
\vspace{-25pt}
\includegraphics[width=0.9\textwidth]{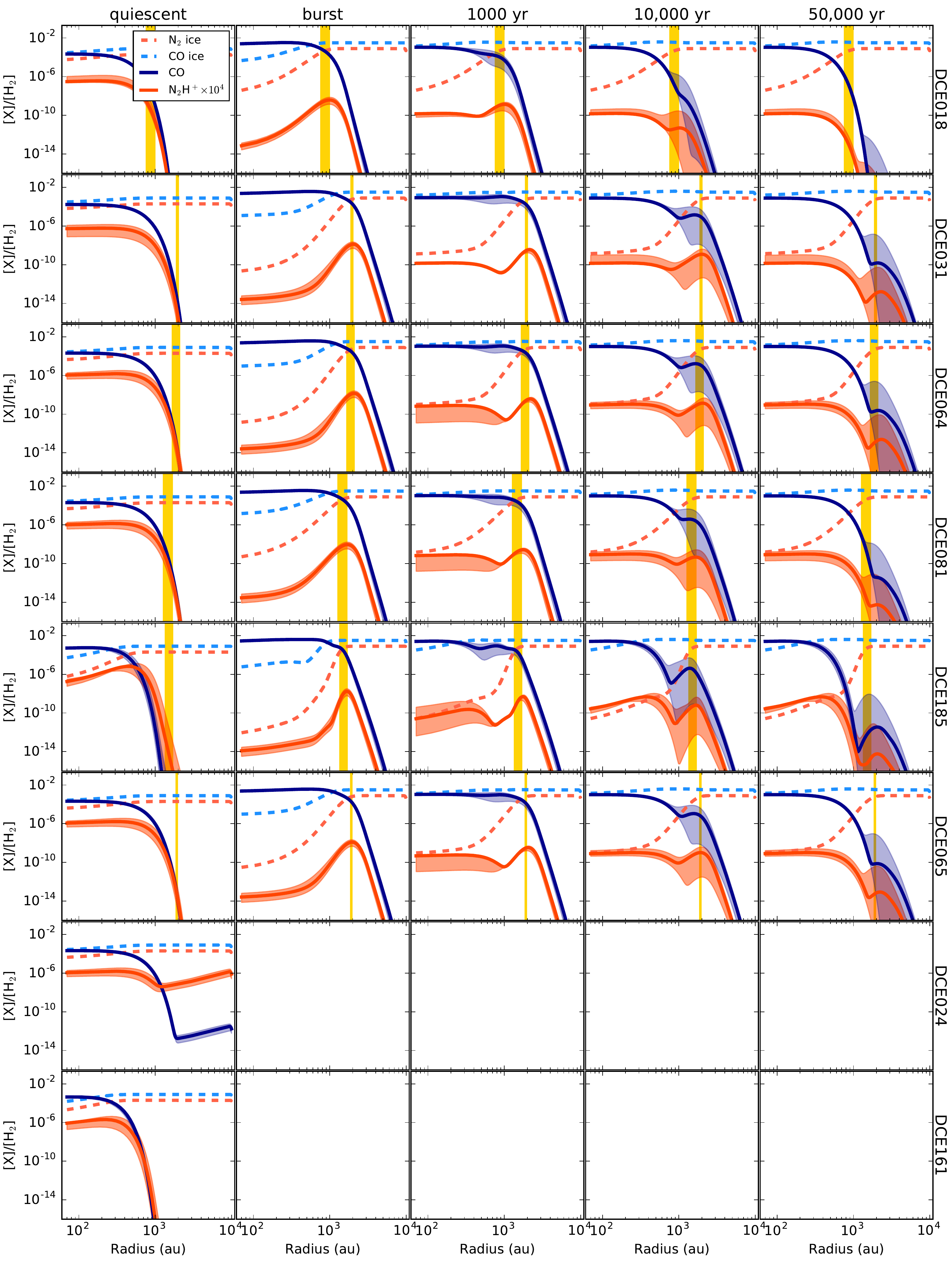}
\caption{Same as Figure \ref{fig:model} but with the synthetic abundance profiles smoothed to the 
angular resolution of the observations.}
\label{fig:model_con}
\end{figure*}

To compare the chemical models with the observations, we smooth the modeled abundance profiles 
in Figure \ref{fig:model} to the observed angular resolution. 
We convolve the profiles with a Gaussian of FWHM equal to the FWHM of the beam (Table \ref{tab:obs}). 
The convolved abundance profiles are shown in Figure \ref{fig:model_con}. 
These profiles indicate that the inner N$_2$H$^+$ peak after the burst cannot be resolved 
in our current data.

The observed N$_2$H$^+$ abundance profile is well described by the models during the quiescent 
or post-burst phases for almost all VeLLOs (Figure \ref{fig:model_con}).
This indeed confirms that five VeLLOs have undergone episodic accretion, because the N$_2$H$^+$ peak could not have moved outward to radii of $\sim$1000~au 
with the current luminosity.  This is only possible through the cloud core being heated up considerably. 
In addition, the time-dependent model suggests that the radius of the N$_2$H$^+$ abundance peak 
does not significantly change in $\sim$10,000 yr. 
This result suggests that the position of the N$_2$H$^+$ peak
can reflect the highest luminosity that the central source reached in the 
past accretion burst.

\section{DISCUSSION}
\label{sec:diss}
\subsection{Occurrences of burst or not?}
\begin{figure}
\includegraphics[width=0.5\textwidth]{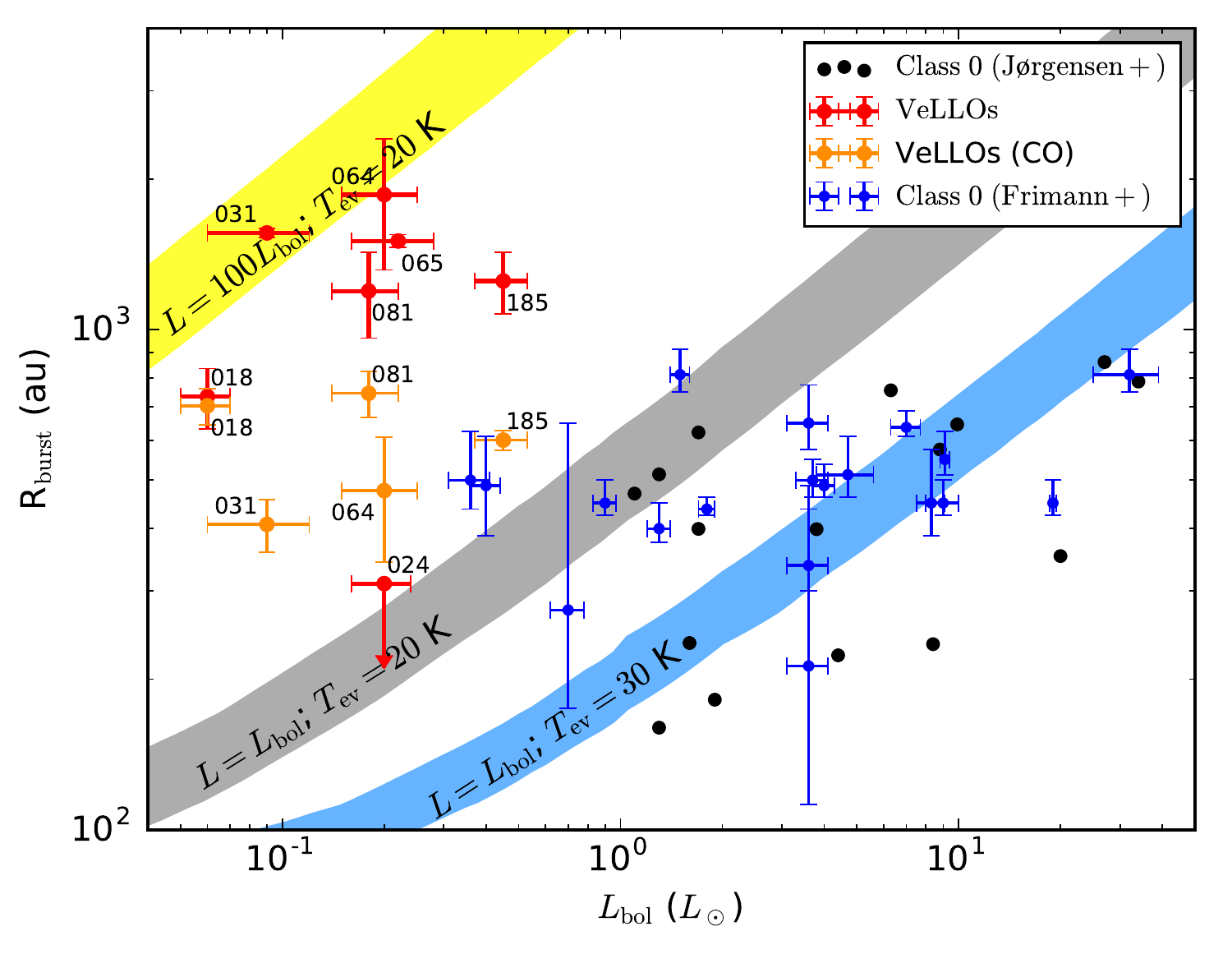}
\caption{Radii of the CO snow line versus the bolometric luminosity (red). 
The radii of the red points correspond to $R_{\rm burst}$ in Table \ref{tab:result}.
The black points and blue points are taken from \citet{jo15} and \citet{fr17}, respectively. 
The radii from the literature are measured through the CO extended emission, and for comparison, 
the orange points indicate the radii calculated through CO observations of VeLLOs.
Considering the envelope mass of $0.5-3.0 M_\odot$, the grey area and blue area indicate the modeled radii 
where the dust temperatures are 20~K and 30~K, respectively, with the bolometric luminosity.
The yellow area shows that for a dust temperature of 20~K and assuming a 100 times higher luminosity.}
\label{fig:L_vs_R}
\end{figure}

Figure \ref{fig:L_vs_R} shows the radius of the CO snow line measured using N$_2$H$^+$ 
as a function of bolometric luminosity, i.e. $R_{\rm burst}$ in Table \ref{tab:result}.
By comparing the location of the CO snow line predicted for the current luminosity to the measured value, 
we find at least five of our targets likely experienced a recent accretion burst.
Excluding DCE161 (see section \ref{sec:n2h}), we now discuss whether or not a previous 
burst has occurred toward the seven VeLLOs (see also Figure \ref{fig:abundance}):
\begin{enumerate}
\item Five of the targets (DCE018, 031, 064, 081, 185) have CO snow lines located at larger radii than 
the predicted values from the model using the current luminosity suggesting that they have experienced 
an accretion burst.
\item DCE024 shows no N$_2$H$^+$ depletion in the intensity/abundance profile with very weak C$^{18}$O and 
$^{13}$CO emission peaks at the center.
Given the roughly constant N$_2$H$^+$ abundance profile, DCE024 is most likely at a quiescent accretion phase 
without a recent burst.
\item The depletion of N$_2$H$^+$ in DCE065 could originate from a previous accretion burst or the depletion 
of the parent molecules N$_2$ by freeze-out onto dust grains (see section \ref{sec:profile}).
\end{enumerate}
As a result, out of the seven VeLLOs, five show evidence for the occurrence of a past burst, one has no 
recent accretion burst, and one is ambiguous.
Statistically, five to six VeLLOs with a past burst out of seven VeLLOs would result in a probability 
of 71\% to 86\% with an uncertainty of 17\% to 13\% assuming binomial statistics.

\tabletypesize{\scriptsize}
\tabcolsep=0.3cm
\begin{deluxetable*}{ccccccc}
\tabletypesize{\tiny}
\tablecaption{Chemical network reactions and adopted rate coefficients}
\tablehead{
ID
& Reaction
& $\zeta$
& $\alpha$
& $\beta$
& $\gamma$
\\
& 
& \colhead{(s$^{-1}$)}
& \colhead{(cm$^{3}$ s$^{-1}$)}
& \colhead{}
& \colhead{(K)}
}
\startdata
1	& H$_2^+$ + H$_2$ $\longrightarrow$ H$_3^+$ + H	& ...	& 2.08$\times10^{-9}$	& 0.00	& 0.0	\\
2	& H$_3^+$ + HD $\longrightarrow$ H$_2$D$^+$ + H$_2$	& ...	& 3.50$\times10^{-10}$	& 0.00	& 0.0	\\
3	& H$_2^+$ + e$^-$ $\longrightarrow$ H + H	& ...	& 1.60$\times10^{-9}$	& -0.43	& 0.0	\\
4	& H$_3^+$ + e$^-$ $\longrightarrow$ H$_2$ + H	& ...	& 6.70$\times10^{-8}$	& -0.52	& 0.0	\\
5	& H$_2$D$^+$ + e$^-$ $\longrightarrow$ H$_2$ + D	& ...	& 6.79$\times10^{-8}$	& -0.52	& 0.0	\\
6	& HCO$^+$ + e$^-$ $\longrightarrow$ CO + H	& ...	& 2.80$\times10^{-7}$	& -0.69	& 0.0	\\
7	& DCO$^+$ + e$^-$ $\longrightarrow$ CO + D	& ...	& 2.40$\times10^{-7}$	& -0.69	& 0.0	\\
8	& N$_2$H$^+$ + e$^-$ $\longrightarrow$ N$_2$ + H	& ...	& 2.60$\times10^{-7}$	& -0.84	& 0.0	\\
9	& N$_2$D$^+$ + e$^-$ $\longrightarrow$ N$_2$ + D	& ...	& 2.60$\times10^{-7}$	& -0.84	& 0.0	\\
10	& H$_3^+$ + CO $\longrightarrow$ HCO$^+$ + H$_2$	& ...	& 1.61$\times10^{-9}$	& 0.00	& 0.0	\\
11	& H$_2$D$^+$ + CO $\longrightarrow$ DCO$^+$ + H$_2$	& ...	& 5.37$\times10^{-10}$	& 0.00	& 0.0	\\
12	& H$_2$D$^+$ + CO $\longrightarrow$ HCO$^+$ + HD	& ...	& 1.07$\times10^{-9}$	& 0.00	& 0.0	\\
13	& N$_2$H$^+$ + CO $\longrightarrow$ HCO$^+$ + N$_2$	& ...	& 8.80$\times10^{-10}$	& 0.00	& 0.0	\\
14	& N$_2$D$^+$ + CO $\longrightarrow$ DCO$^+$ + N$_2$	& ...	& 8.80$\times10^{-10}$	& 0.00	& 0.0	\\
15	& H$_3^+$ + N$_2$ $\longrightarrow$ N$_2$H$^+$ + H$_2$	& ...	& 1.80$\times10^{-9}$	& 0.00	& 0.0	\\
16	& H$_2$D$^+$ + N$_2$ $\longrightarrow$ N$_2$D$^+$ + H$_2$	& ...	& 6.00$\times10^{-10}$	& 0.00	& 0.0	\\
17	& H$_2$D$^+$ + N$_2$ $\longrightarrow$ N$_2$H$^+$ + HD	& ...	& 1.20$\times10^{-9}$	& 0.00	& 0.0	\\
18	& HCO$^+$ + D $\longrightarrow$ DCO$^+$ + H	& ...	& 1.00$\times10^{-9}$	& 0.00	& 0.0	\\
19	& DCO$^+$ + H $\longrightarrow$ HCO$^+$ + D	& ...	& 2.20$\times10^{-9}$	& 0.00	& 796.0	\\
20	& H$_3^+$ + D $\longrightarrow$ H$_2$D$^+$ + H	& ...	& 1.00$\times10^{-9}$	& 0.00	& 0.0	\\
21	& H$_2$D$^+$ + H $\longrightarrow$ H$_3^+$ + D	& ...	& 2.00$\times10^{-9}$	& 0.00	& 632.0	\\
22	& N$_2$H$^+$ + D $\longrightarrow$ N$_2$D$^+$ + H	& ...	& 1.00$\times10^{-9}$	& 0.00	& 0.0	\\
23	& N$_2$D$^+$ + H $\longrightarrow$ N$_2$H$^+$ + D	& ...	& 2.20$\times10^{-9}$	& 0.00	& 550.0	\\
24	& H$_2$ + cr $\longrightarrow$ H$_2^+$ + e$^-$	& 1.21E-17	& ...	& ...	& ...	\\
25	& H$_2$D$^+$ + H$_2$ $\longrightarrow$ H$_3^+$ + HD	& ...	& 1.40$\times10^{-10}$	& 0.00	& 232.0	\\
26	& H$_2$D$^+$ + H$_2$ $\longrightarrow$ H$_3^+$ + HD	& ...	& 7.00$\times10^{-11}$	& 0.00	& 61.5
\enddata
\label{tab:chemistry}
\end{deluxetable*}

It is noteworthy that in a low-density region molecules may not have had enough time to freeze out onto 
dust grains since the assembly of the molecular cloud.
Thus, the radius of the CO snow line may highlight not only the area heated by a luminosity burst but 
also a region where CO has not yet frozen out \citep{jo05}. 
The slow freeze-out process of CO can be seen in the chemical model at a large radius of 
$\gtrsim$2000-3000~au (Figure \ref{fig:model}). 
However, at a radius of $\lesssim$1000-2000~au, CO can freeze out within $\sim$10,000~yr
(Figure \ref{fig:model}) which is much shorter than the lifetime of a starless core of $\sim 10^{5}$~yr \citep{en08}. 
Therefore, CO within the mentioned radius is most likely evaporated by a past luminosity outburst.

Figure \ref{fig:L_vs_R} also shows the data points from \citet{jo15} and \citet{fr17} which plot the 
half width at half maximum (HWHM) of the C$^{18}$O (2--1) emission as a function of the bolometric luminosity. 
For comparison, we plot in orange the HWHM of the CO isotopologue emission from our targets (Table \ref{tab:gauss}). 
This comparison indicates that the radii of the CO snow lines are larger than the HWHM of the CO isotopologue 
emission by a factor ranging from $1-4$, while the model in \citet{an16} found a factor of 1.6. 
The scatter in the observed HWHM likely originates from the spatial resolution of observations (Table \ref{tab:obs}). 
In addition, the outflow contamination can not be completely removed from the CO isotopologue observations.

\subsection{Mass accretion rate during burst}
\label{sec:macc}
The estimated $L_{\rm burst}$ (Table \ref{tab:result}) provides us with an opportunity to estimate the 
mass accretion rate during the accretion burst.
Assuming $L_{\rm acc}=L_{\rm burst}$, we estimate the mass accretion rate ($\dot{M}_{\rm acc}$) during the 
burst phase using $L_{\rm acc}=\frac{GM_{\rm star}\dot{M}_{acc}}{R}$ where G is the gravitational constant, 
$R$ is the protostellar radius (assumed to be 3~$R_\odot$; \citealt{du10a}) and $M_{\rm star}$ is the mass of 
the central source (assumed to be 40~$M_{\rm Jupiter}$; \citealt{hu06}).
As a result, we find mass accretion rates during the burst phase of 
$\sim6\times10^{-6}-4\times10^{-5}$~$M_\odot$ yr$^{-1}$ with a median of $\sim2.0\times10^{-5}$ $M_\odot$ yr$^{-1}$.
We note that this estimate could be considered to be a lower limit because 
(1) a fraction of N$_2$H$^+$ may have recovered since the past burst such that the measured 
N$_2$H$^+$ depletion radius is smaller than that during the accretion burst, and (2) the protostellar disk or torus-like structure, if present, can shield the outer 
envelope from heating by the central protostars \citep{mu15,ra17}.  This latter point is important to consider given that the abundance 
profiles are derived from the cuts perpendicular to the outflows.

Given the derived mass accretion rate, we can further estimate how much time a protostar spends in the burst phase.
Assuming that most of the material is accreted during the burst at an accretion rate of 
$\sim2.0\times10^{-5}$ $M_\odot$ yr$^{-1}$, the total time of the burst phase is estimated to be $\sim$25,000~yr which is the time necessary to acquire the 
typical stellar mass of 0.5~$M_\odot$.
Taking the lifetime of the embedded phase (Class 0/I) to be $0.4-0.5$~Myr \citep{ev09,du14,ca16}, 
we find that a protostar spends $\sim 5-6$\% of the time in the burst phase, which is on the same order 
as the simulation results of $\sim1.3$\% from \citet{du12} and is also consistent with the observational 
statistical results of $\sim$5\% with $\dot{M}\gtrsim10^{-5} M_\odot$ yr$^{-1}$ from \citet{en09}.

\subsection{Timeline of the episodic accretion process}
One of the most important parameters in the episodic accretion process is the time interval between 
two accretion bursts because this timescale moderates the effect of radiative feedback on disk fragmentation \citep{me16}. Disk fragments may later be ejected forming a proto-brown dwarf \citep{re01,ba02,ri03,st09,ba12}, or fall onto the central star causing an accretion burst \citep{vo05,vo10,du12}, or migrate to a stable orbit 
resulting in a multiple system. Multiplicity is discussed further in section \ref{sec:multiplicity}.
Out of our seven VeLLOs, we found that at least five show evidence for the occurrence of a past burst.
This fraction is similar to or slightly higher than the value of $20-50$\% found by 
\citet{jo15} and \citet{fr17} in Class 0/I sources based on SMA C$^{18}$O observations.
Taking the time scale of 10,000~yr for CO to refreeze out \citep{vi12,vi15}, we suggest $71 \pm 17\%$ to $86 \pm 13\%$ 
of these VeLLOs have experienced a past accretion burst within 10,000~yr. 
If we assume that all protostars are undergoing episodic accretion, this result suggests a timescale 
of intervals between bursts of $12,000 - 14,000$~yr in VeLLOs, in comparison to $20,000 - 50,000$~yr 
from \citet{jo15} and \citet{fr16}. 
However, if an accretion burst lasts for 100 to 200~yr \citep{vo05} and in turn accretes material with 
a mass of $\sim 2-4 \times10^{-3}$~$M_\odot$ (section \ref{sec:macc}), it would take $125-250$ bursts, 
i.e.\ $1.5-3.5\times10^6$ yr in total to build a star with a mass of 0.5 $M_\odot$.  This is too long compared with the lifetime of the Class 0/I stage, $0.4-0.5$~Myr \citep{ev09,du14,ca16}. 
Thus, either a significant fraction of mass ($\gtrsim$70\%) accretes during the quiescent phase, or stronger 
outbursts occur at a later evolutionary stage assuming VeLLOs are early Class 0 sources \citep{hs15,hs17}. 
Alternatively, these VeLLOs will form very low-mass stars with a mass $\ll 0.5$~$M_\odot$.
Another possible explanation is that the interval between bursts is in fact shorter than the time scale 
for CO to freeze out (10,000~yr). For such a case, this means that six out of seven sources are undergoing episodic accretion whereas the remaining source, DCE024, has not yet started this process.

In addition to the time interval between bursts, the detection of accretion bursts in VeLLOs suggests that 
the episodic accretion process may start from a very early stage since VeLLOs could be extremely young 
protostars \citep{du14,hs15,hs17}.
The onset time of episodic accretion is important because it is associated with the origin of accretion bursts.
\citet{vo05,vo10} suggested that dense protostellar clumps fall onto the central protostar, causing an accretion burst. Further, \citet{vo10,vo13} found that this dense clump may form through disk fragmentation under the conditions 
of a low Toomre Q-parameter \citep{to64} and a short local cooling time.
However, this process requires a fairly massive disk around the central star \citep{vo13,kr16}. The presence (or otherwise) of disks in VeLLOs remains unclear because they are considered to be 
either very early-stage protostars or extremely low-mass protostars \citep{du14}.
Though the presence of a protostellar outflow strongly suggests the existence of a disk, the resolution of our data 
($\sim300-600$~au) is insufficient to resolve it for most of the targets.
Although protostellar disks have been detected in some Class 0 sources \citep{le11,to12b,mu13a,mu13b,oh14}, 
the disks in VeLLOs are likely still forming if they are indeed extremely young Class 0 sources.
For example, \citet{ye15,ye17} suggest that the disk size is very small (6~au) in one of our targets, DCE185 (IRAS 16253) 
while it shows a clear evidence of a recent accretion burst.
As a result, the high fraction of burst signatures in VeLLOs is in conflict with the simulation results \citep{vo13,vo15} 
in which episodic accretion processes are found to occur preferentially during the Class I stage after the formation of 
a star-disk system.
To address this issue, we propose two possible scenarios:
(1) the accretion burst may not necessarily be triggered by an infalling fragment but by a different mechanism regulating 
the accretion process, or 
(2) the fragment may not necessarily be formed in a disk (disk fragmentation) but in a large-scale envelope or 
pseudo-disk like structure.
It should also be noted that the disk can be consumed by past accretion bursts \citep{jo15}. 
However, in such a case, these VeLLOs would have once contained a massive disk before the burst.

\subsection{Multiplicity}
\label{sec:multiplicity}
Multiplicity is common in embedded protostellar systems, and fragmentation of the cloud core or disk is 
widely agreed to be the mechanism of formation.
Numerical calculations have shown that the thermal feedback from newly formed protostars can suppress 
fragmentation in cloud cores \citep{kr14}, while episodic accretion can moderate the effect of thermal feedback. During the interval of an accretion burst, the cloud core can cool down and fragmentation can occur 
in the disk \citep{st12} or in the rest of the cloud core where the temperatures fall well below 100~K.
On the other hand, the mass of the core may be an important factor in fragmentation as well \citep{of16}.
VeLLOs which are undergoing episodic accretion would be expected to be multiple protostellar systems.
However, none of the VeLLOs studied so far have shown convincing evidence of multiplicity, 
with probable companions only detected in the submillimeter but not in the infrared \citep{ta13,ch12} or 
through outflow kinematics \citep{hs16}.
At separations larger than 250~au, the sources in our sample do not show a secondary companion nor signs of multiplicity. 
DCE065 may be the exception, with the SMA continuum observations hinting at a possible binary system; 
however, \citet{to16} did not find indications of companion sources in DCE065.
Further studies would be needed to determine whether DCE065 is a true proto-binary.
The angular resolution of our observations, $\sim1\farcs4-3\farcs1$, is not enough to probe the existence 
of a companion at separations smaller than 250~au.
A recent unbiased survey at millimeter wavelengths of all protostars in the Perseus molecular cloud 
probed down to 15~au separations \citep{to16}. 
The Perseus VeLLOs included in our sample (DCE064, 065 and 081) have been found to be single protostars.
The measured $L_{\rm bol}$ of an unresolved binary is higher than that of a single protostar, even if 
the protostars are at similar evolutionary stages \citep{mu16}. 
If we compare the $L_{\rm bol}$ from the VeLLOs in Perseus to the rest of our sample, the luminosities are similar. 
Thus, either the rest of the sample is also composed of single protostars or they are binaries with even 
dimmer components.

Episodic accretion cannot be a hindering factor in fragmentation, since several embedded multiple protostellar 
systems in Perseus have evidence of accretion bursts \citep{fr17}.
The low frequency of multiplicity in our sample of VeLLOs could be explained by a few possible scenarios. One possibility is that the time between accretion bursts is too short to allow the gas to cool down 
enough for fragmentation to take place.
Even if there were areas of cold gas, there may not be areas of high enough density that can trigger 
instabilities and fragment.
Another possibility is that, while temperature may play an important role in the star formation process, 
it may not be the deciding factor in the formation of multiple protostellar systems through fragmentation.

\section{SUMMARY}
\label{sec:sum}
We present an ALMA survey of N$_2$H$^+$ (1--0), $^{13}$CO (1--0), C$^{18}$O (1--0), C$^{17}$O (1--0) line emission 
and dust continuum emission in eight VeLLOs. 
We use the molecular pair, N$_2$H$^+$ and CO isotopologues, to probe the positions of the CO snow lines 
in the natal envelopes. 
We use the measured position of the CO snow line, together with that predicted from the current luminosity, 
to identify VeLLOs that have experienced a recent accretion burst. 
We summarize our results as follows.

\begin{enumerate}
\item We found five to six VeLLOs out of seven that have experienced recent accretion bursts. This fraction is larger than that found for Class 0/I objects (\citealp[$\sim20-50$\%,][]{jo15,fr17}), 
and it implies a time interval between accretion burst of $12,000-14,000$~yr, assuming a CO refreeze-out time of 10,000~yr.
\item Our chemical model reproduces well the observed N$_2$H$^+$ depletion toward the source center due to a 
past accretion burst. 
In addition, we find that the radius of the CO snow line remains relatively static after the burst, 
suggesting that it can be used as an indicator of the burst luminosity.
\item From the observed positions of the CO snow lines, we estimate mass accretion rates of 
$\sim 6\times10^{-6}-4\times10^{-5}$~$M_\odot$~yr$^{-1}$ with a median of $\sim2.0\times10^{-5}$~$M_\odot$ yr$^{-1}$ 
during the burst. 
Given the lifetime of the protostellar embedded phase, we suggest that a protostar spends $\sim 5-6$\% of the time 
in the burst phase to form a typical mass (0.5 $M_\odot$) star.
\item The evidence of accretion bursts in VeLLOs is in conflict with simulations \citep{vo15} which suggest that 
episodic accretion should preferentially occur during the Class I stage after accreting sufficient material onto a protostellar disk. 
We propose two possibilities to solve this issue: 
(1) the accretion burst may not necessarily be triggered by an infalling fragment but by a different mechanism 
regulating the accretion process, or 
(2) the fragment may not necessarily be the product of disk fragmentation, but form in a large-scale envelope or 
pseudo-disk structure.
\item In our sample of VeLLOs, we find no evidence for multiplicity down to spatial scales of 250~au, 
implying a low frequency of multiplicity.
We suggest that 
(1) the short time interval between bursts prevents the gas from cooling down and fragmenting to form a binary system, and/or
(2) given the cold envelopes around VeLLOs, the absence of multiplicity suggests that temperature may not be the deciding 
factor for fragmentation.
\end{enumerate}

The authors thank Prof. Jes. K. J\o rgensen and Prof. Yuri Aikawa for providing valuable discussions.
We thank Dr. Chao-Ling Hung for sharing the SMA data in Figure \ref{fig:profile}, \ref{fig:abundance}, and \ref{fig:DCE065}.
We express our gratitude to the anonymous referee for the constructive comments that improved the clarity of this paper.
We are thankful for the help from ALMA Regional Center in Taiwan.
This paper makes use of the following ALMA data: ADS/JAO.ALMA\#2015.1.01576.S. ALMA is a partnership of ESO 
(representing its member states), NSF (USA) and NINS (Japan), together with NRC (Canada), NSC and ASIAA (Taiwan), 
and KASI (Republic of Korea), in cooperation with the Republic of Chile. 
The Joint ALMA Observatory is operated by ESO, AUI/NRAO and NAOJ. 
The 2015.1.01576.S data was obtained by T.H.H. while he was a PhD student at National Tsing Hua University, 
Taiwan, under the supervision of S.P.L. 
T.H.H. and S.P.L. are thankful for the support of the Ministry of Science and Technology (MoST) of Taiwan 
through Grants 102-2119-M-007-004-MY3, 105-2119-M-007-024, and 106-2119-M-007-021-MY3. 
N.H. and T.H.H. acknowledge a grant from MoST 106-2112-M-001-010 in support of this work.
C.W. acknowledges financial support from the University of Leeds.

\bibliographystyle{aa} 
\bibliography{references.bib} 

\newpage

\appendix
\setcounter{figure}{0}
\setcounter{table}{0}
\renewcommand{\thefigure}{A\arabic{figure}}

\section{Hyperfine fitting of N$_2$H$^+$ spectra}
\label{sec:tau}
To derive the optical depth, we fit the hyperfine structure for each pixel (0\farcs4) in the N$_2$H$^+$ maps.
The fitting is done for the pixels with S/N $>5$ (the rms noise levels range from 0.21 to 0.58~K at a channel 
width of 0.05 km s$^{-1}$).
For DCE031, the spectra show a ``wing-like'' structure toward the four spikes seen in the images 
(e.g., point 1 in Figure \ref{fig:DCE031}), which are likely associated with the outflow entrained gas. 
We thus fit the spectra only on the central flattened envelope in DCE031.

With the high spectral resolution of $\sim$0.1 km~s$^{-1}$, we find that many of the regions in fact contain 
two (or even more) velocity components along the line of sight (e.g. point 3 in Figure \ref{fig:DCE081}).
For these regions, we fit the spectra with a two-component hyperfine structure.
However, depending on linewidth and separation of the two components, the double-peak structures are sometimes 
only seen in a few of the seven components (e.g. point 4 in Figure \ref{fig:DCE064}). 
In addition, the weak component is usually marginally detected, together with hyperfine anomalies, 
making the fitting with two components more difficult.
As a result, we suggest that the fitting results are decent for the primary component, and for the weak component, 
the centroid velocity and linewidth are broadly reliable since they are less affected by the signal to noise ratio 
and hyperfine anomalies.
Further details on the dynamics are outside the scope of this paper and will be discussed elsewhere.

\begin{figure*}
\includegraphics[width=\textwidth]{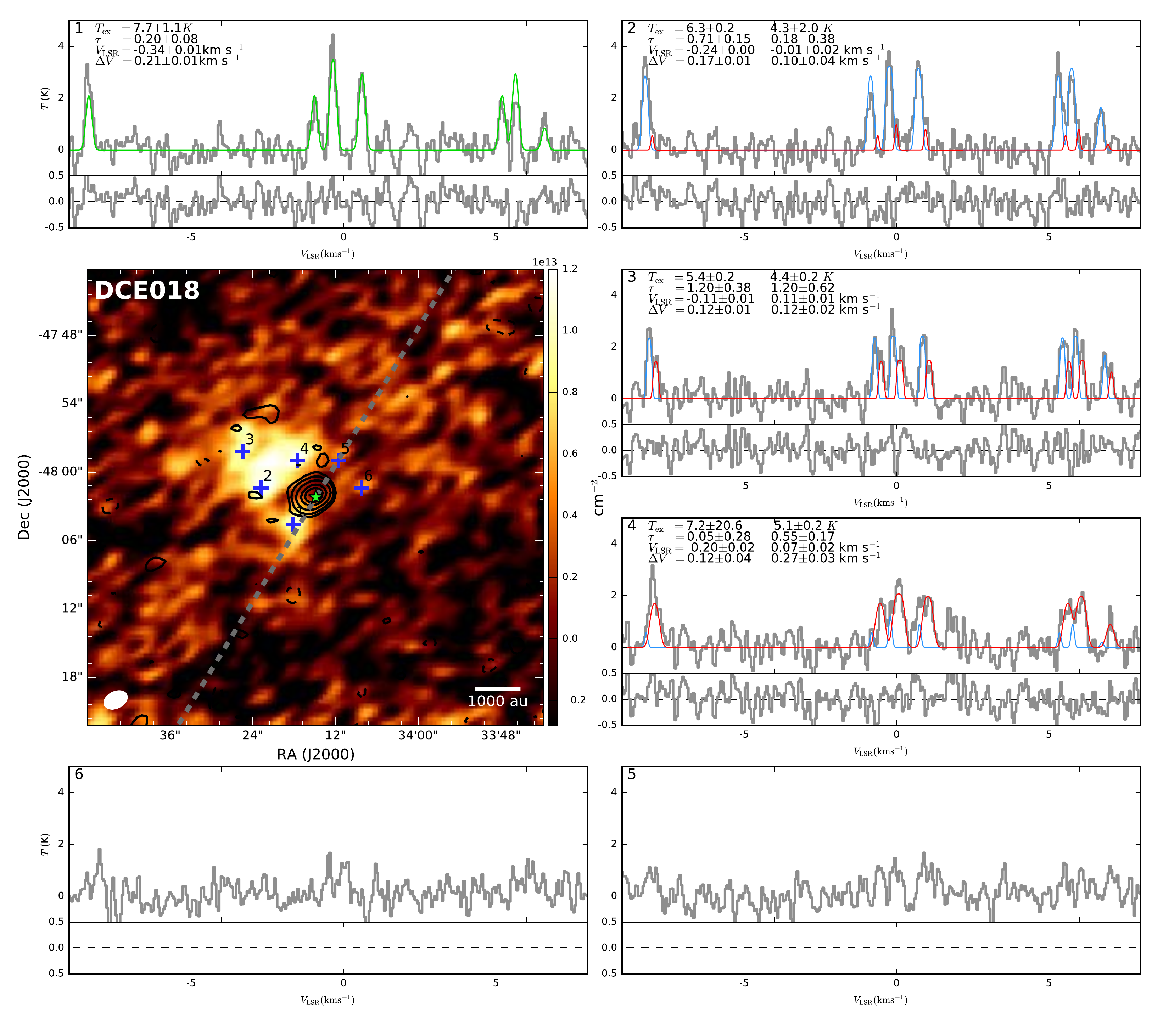}
\caption{N$_2$H$^+$ column density map and spectra of 6 arbitrary selected positions (blue plus signs) in DCE018. 
The green star indicates the infrared source position. 
The black contours show the ALMA 3~mm continuum emission with levels of 3, 5, 10, 20, 30, 50$\sigma$. 
The grey dashed line indicates the cut that is used to measure the intensity profiles (Fig. \ref{fig:profile}) 
and abundance profiles (Fig. \ref{fig:abundance}). 
In each panel for spectra, the best-fit results are shown in green lines for one-component fitting 
and blue and red lines for two-component fitting with the bottom subpanel showing the residuals. 
The best-fit parameters ($\tau$ for the weakest component N$_2$H$^+$ 110--011) are shown in the upper 
left corner in each spectrum panel for blue-shifted (left) and red-shifted (right) components if present.}
\label{fig:DCE018}
\end{figure*}

\begin{figure*}
\includegraphics[width=\textwidth]{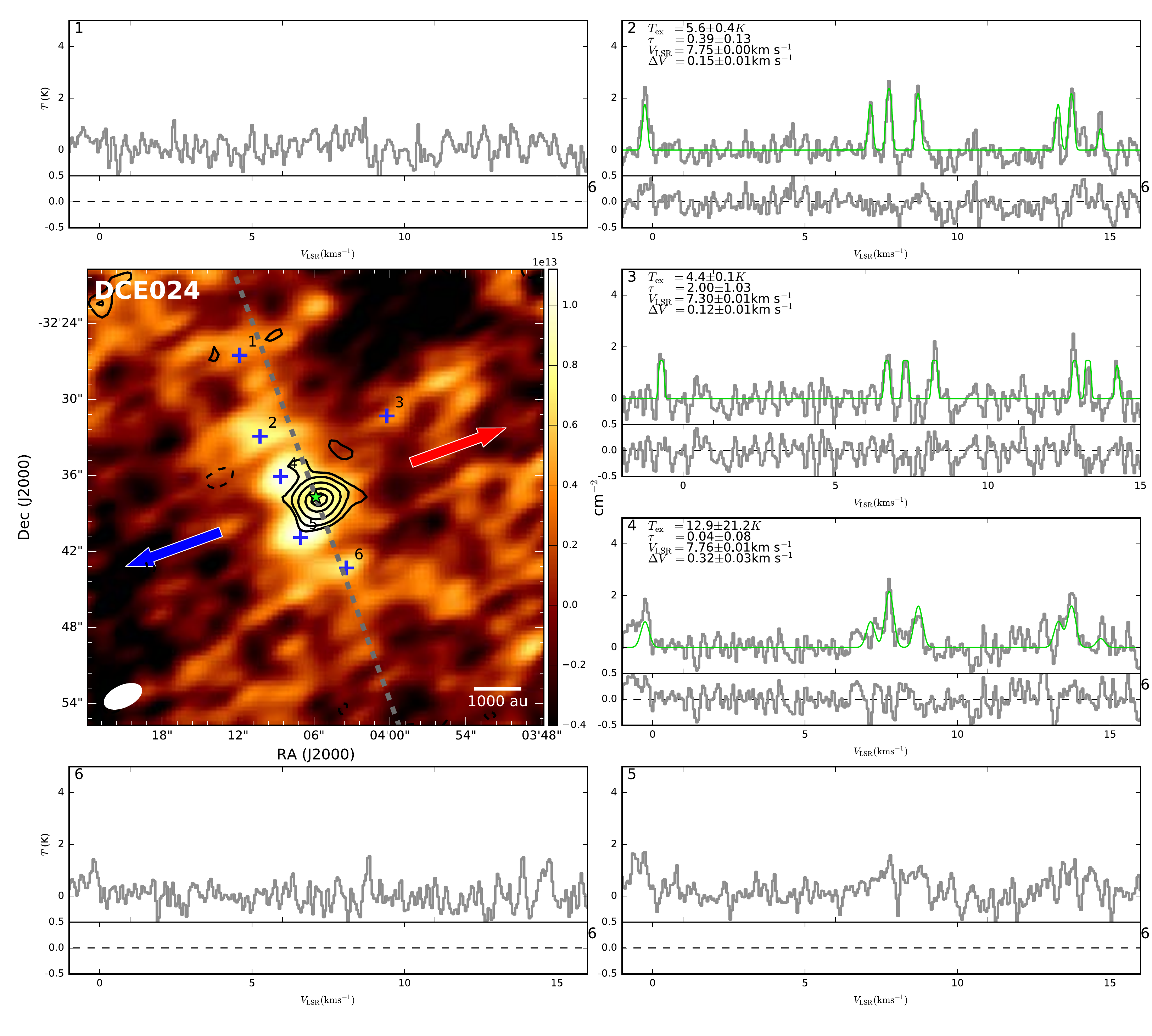}
\caption{Same as Figure \ref{fig:DCE018} but for DCE024.}
\label{fig:DCE024}
\end{figure*}

\begin{figure*}
\includegraphics[width=\textwidth]{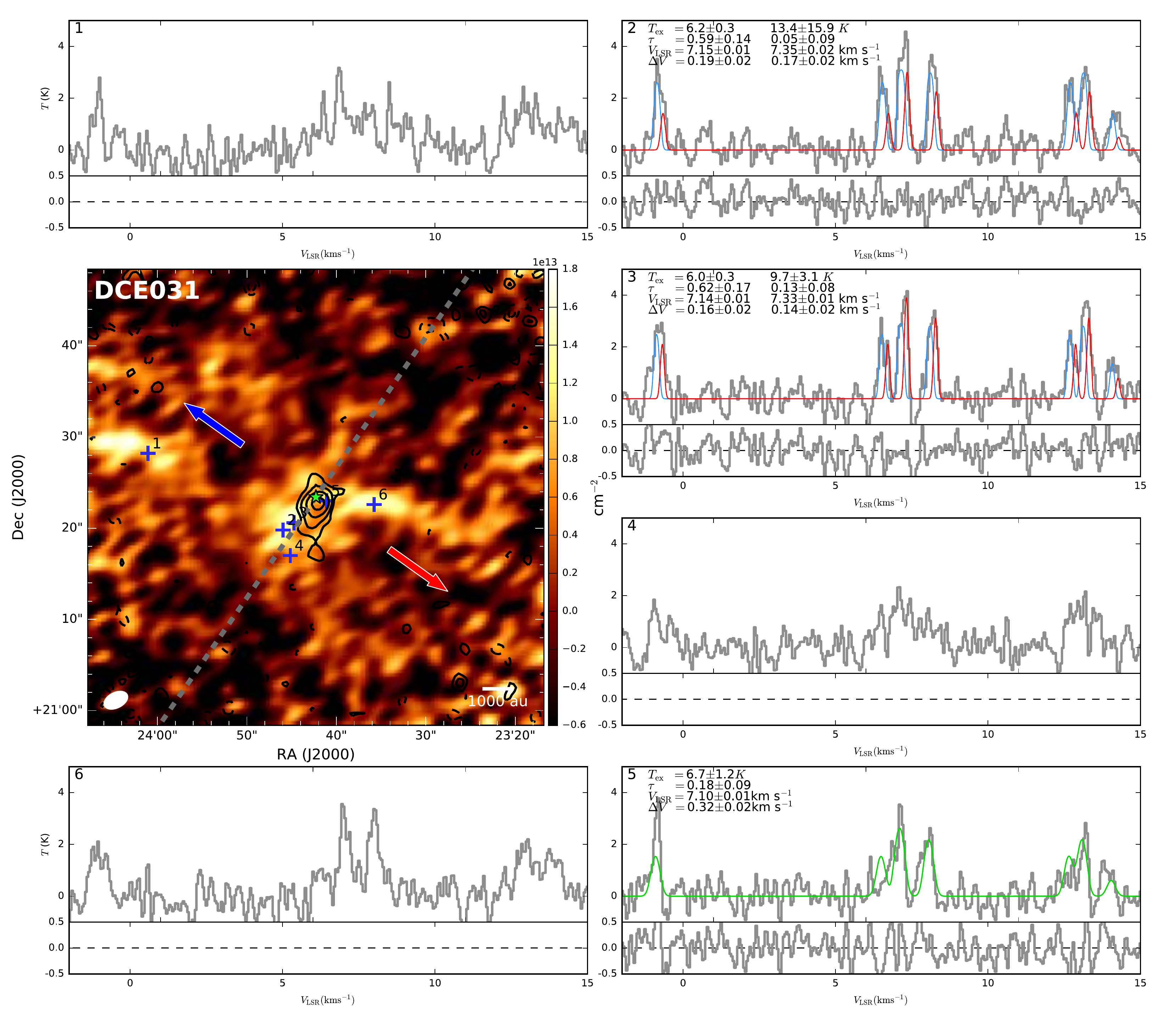}
\caption{Same as Figure \ref{fig:DCE018} but for DCE031.}
\label{fig:DCE031}
\end{figure*}

\begin{figure*}
\includegraphics[width=\textwidth]{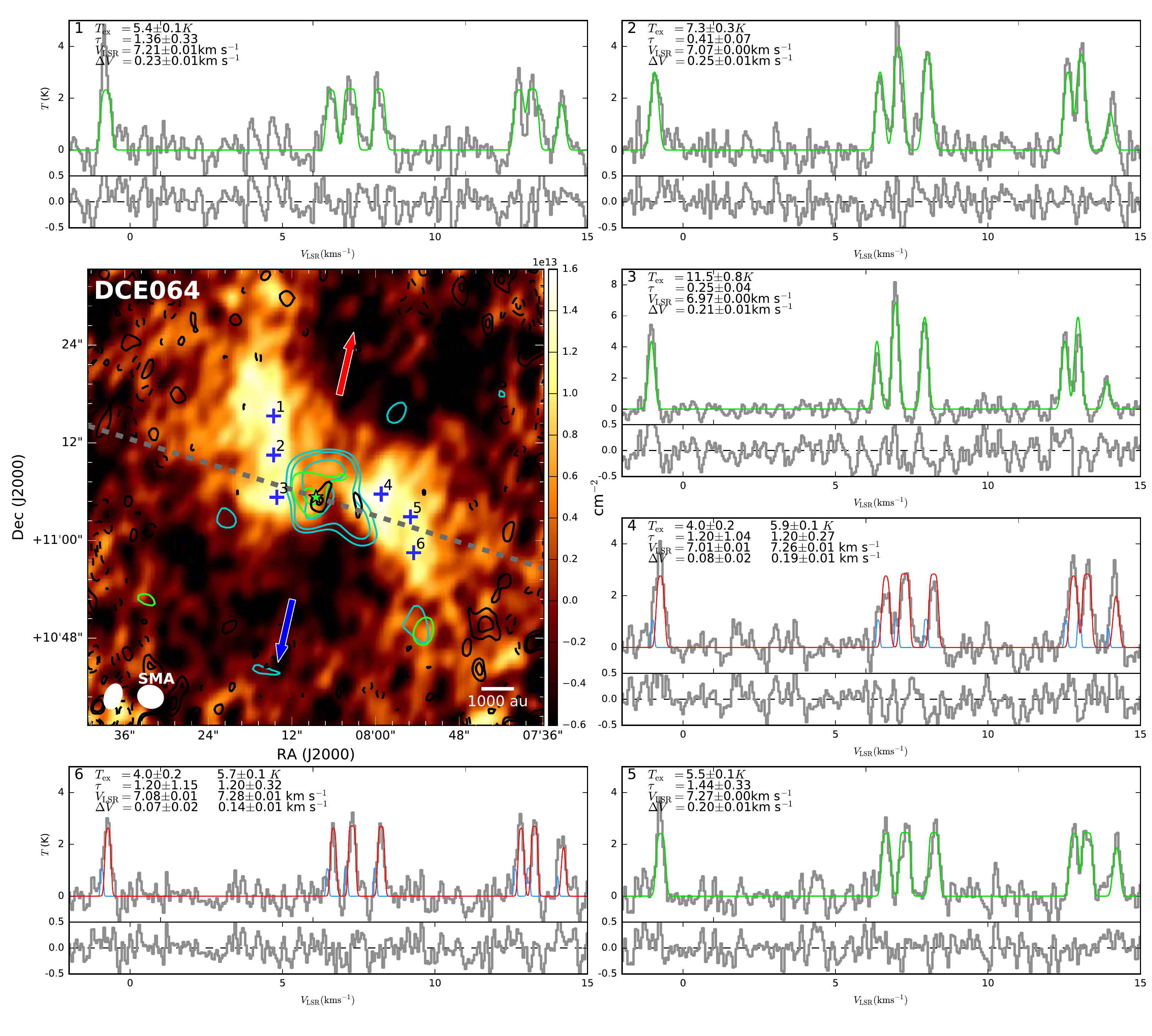}
\caption{Same as Figure \ref{fig:DCE018} but for DCE064. Additional SMA maps are shown in cyan contours 
($^{13}$CO J=2--1) and green contours (C$^{18}$O J=2--1) with the contour levels of 3, 5, 10 $\sigma$. 
The rms noise levels are 0.14 Jy beam$^{-1}$ km s$^{-1}$ with a velocity interval of integration of 
5.2 km s$^{-1}$ to 9.3 km s$^{-1}$ for the $^{13}$CO (2--1) map and 0.08 Jy beam$^{-1}$ km s$^{-1}$ 
with a velocity interval of integration of 6.2 km s$^{-1}$ to 8.9 km s$^{-1}$ for the C$^{18}$O (2--1) map.}
\label{fig:DCE064}
\end{figure*}

\begin{figure*}
\includegraphics[width=\textwidth]{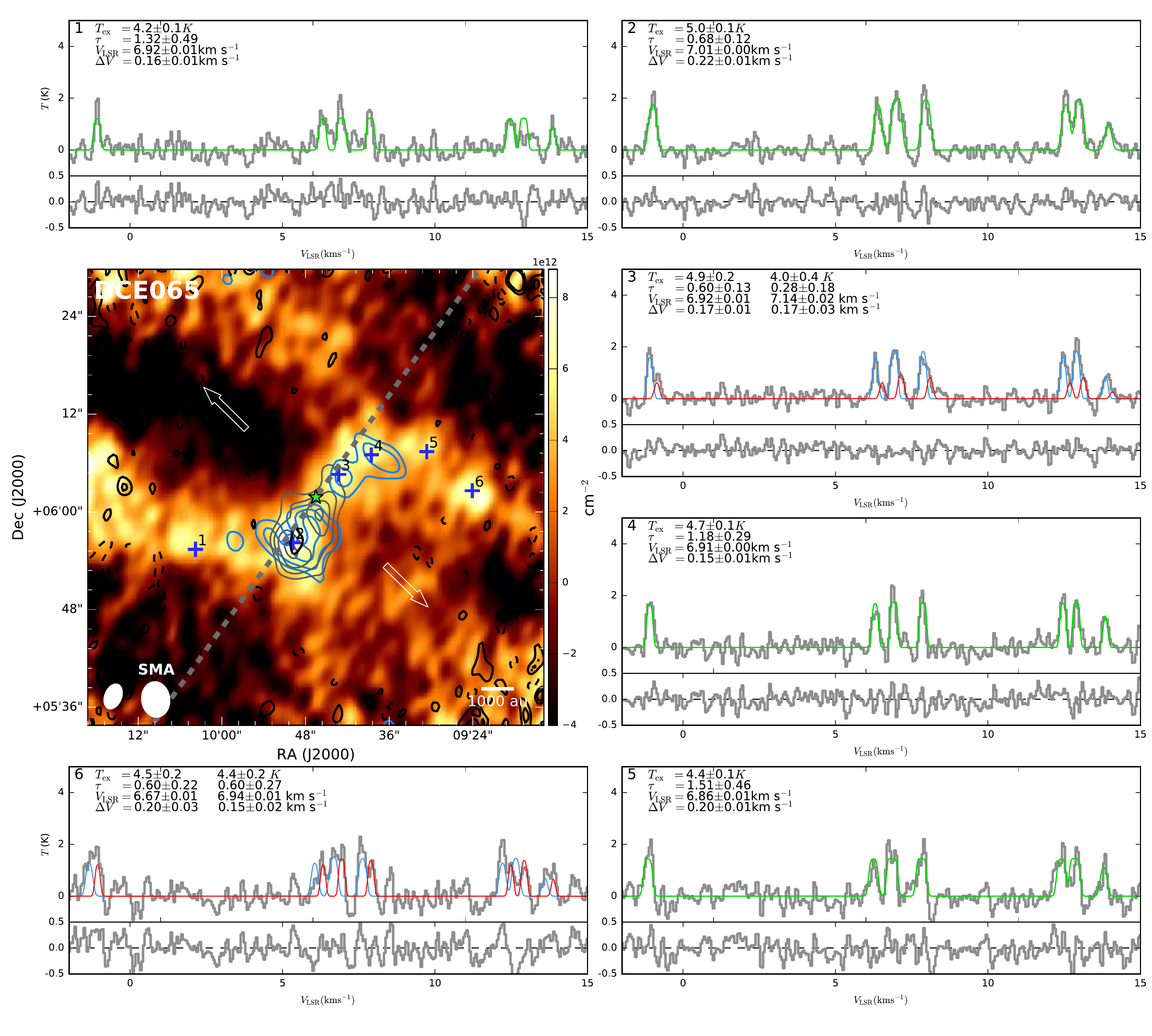}
\caption{
Same as Figure \ref{fig:DCE018} but for DCE065. The gray contours and blue contours show the SMA 1.3~mm 
continuum emission and N$_2$D$^+$ (3--2) emission integrated from 6.2 km s$^{-1}$ to 7.2 km s$^{-1}$, respectively. 
These SMA contour levels start at 3 $\sigma$ with a step of 1 $\sigma$, where $\sigma$ is 0.5 mJy beam$^{-1}$ for 
the continuum map and 0.07 Jy beam$^{-1}$ km s$^{-1}$ for the N$_2$D$^+$ (3--2) map.
}
\label{fig:DCE065}
\end{figure*}

\begin{figure*}
\includegraphics[width=\textwidth]{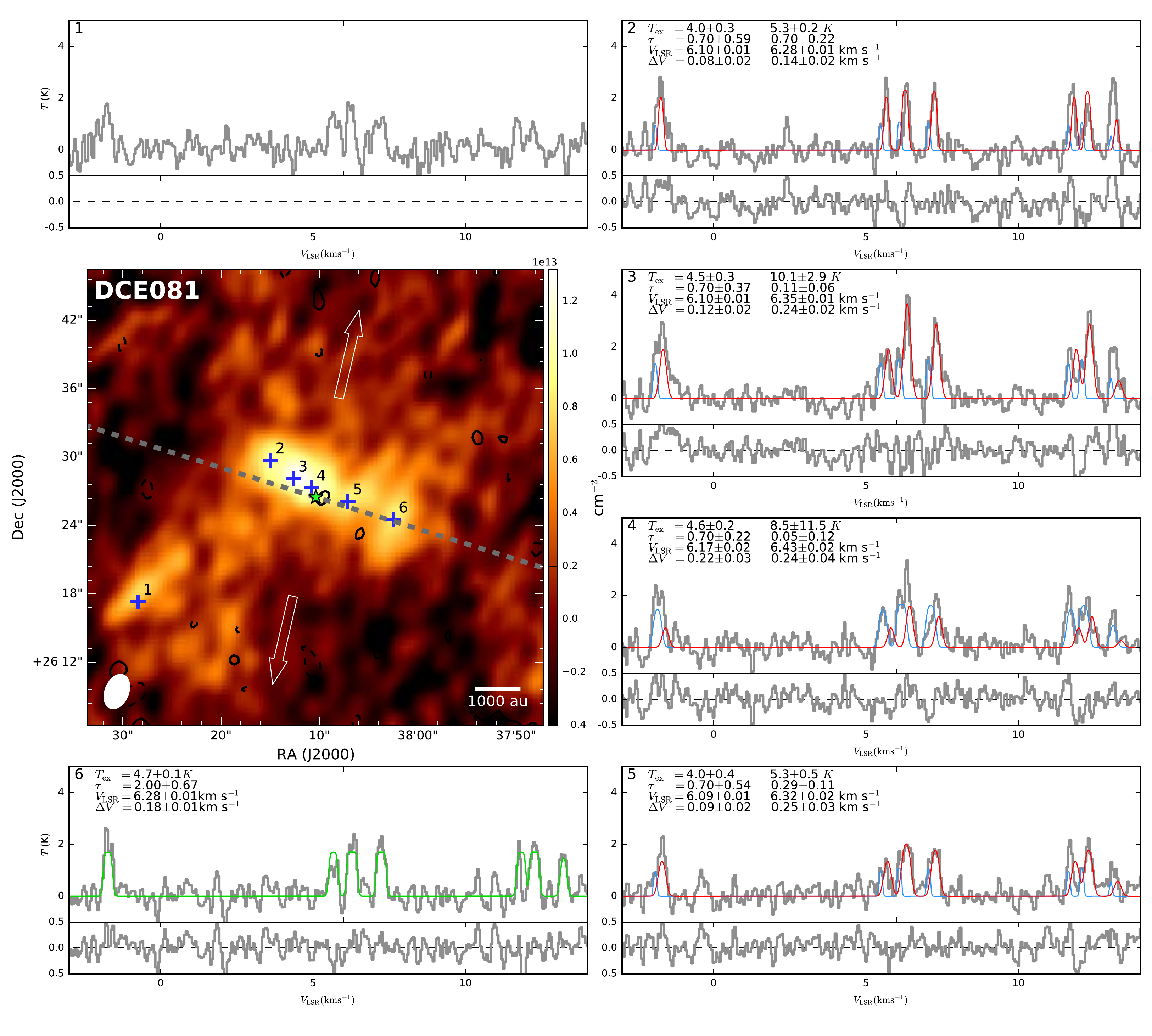}
\caption{Same as Figure \ref{fig:DCE018} but for DCE081.}
\label{fig:DCE081}
\end{figure*}

\begin{figure*}
\includegraphics[width=\textwidth]{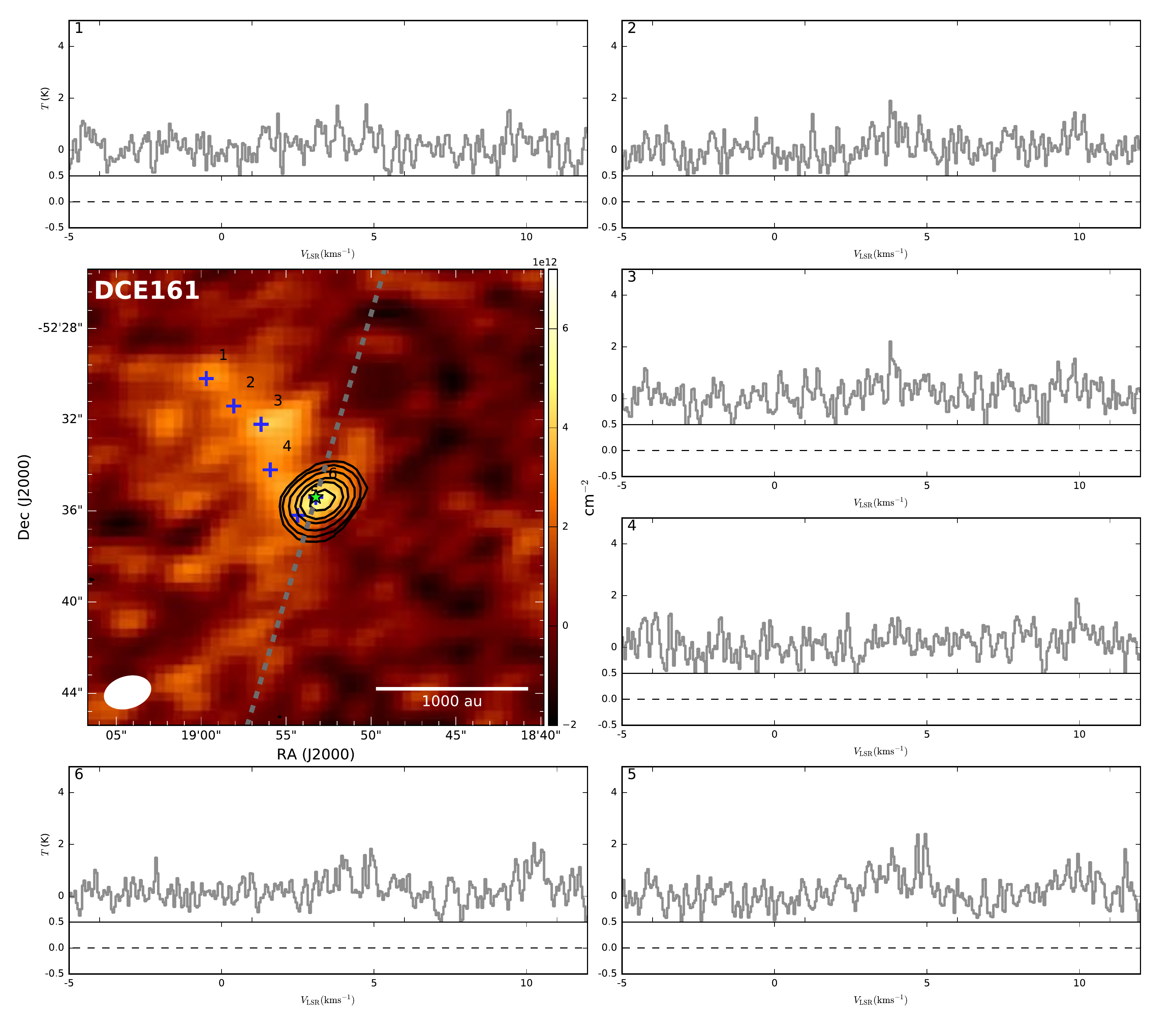}
\caption{Same as Figure \ref{fig:DCE018} but for DCE161. 
The column density was estimated by assuming that all components are optically thin.}
\label{fig:DCE161}
\end{figure*}

\begin{figure*}
\includegraphics[width=\textwidth]{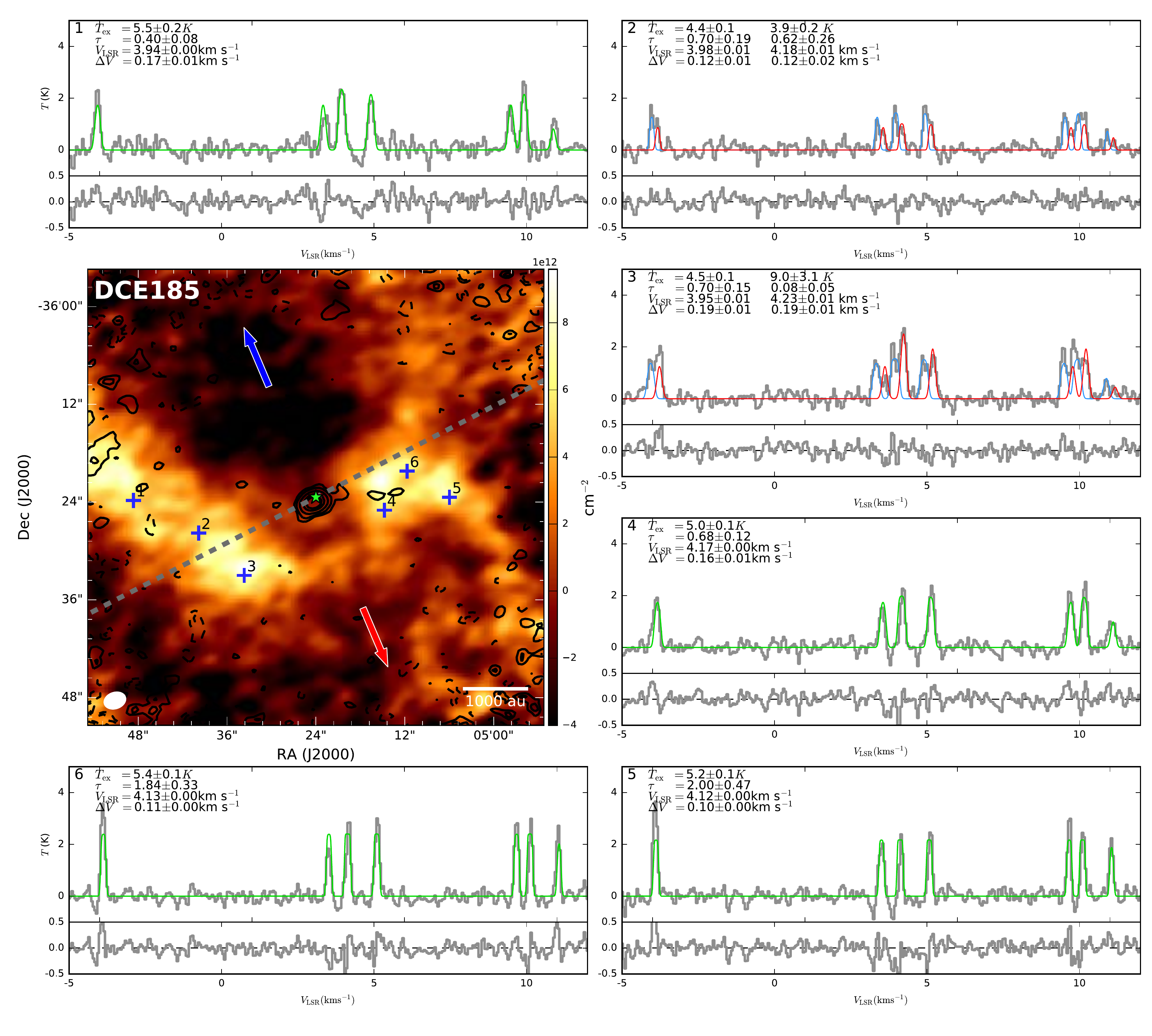}
\caption{Same as Figure \ref{fig:DCE018} but for DCE185.}
\label{fig:DCE185}
\end{figure*}

The results of hyperfine fitting are shown in Figures \ref{fig:DCE018} to \ref{fig:DCE185} (note that the optical 
depth is for the weakest component 110--011 which is one third of the isolated component 101--012). 
When hyperfine anomalies, with reversed relative intensities, are not present, the optical depth is generally low 
($\lesssim 0.5-0.7$) with the one- or two-component fitting in the primary component, which are relatively reliable. 
Note that we plot the N$_2$H$^+$ spectra toward the position with a relatively high integrated intensity where the 
column densities as well as $\tau$ are expected to be high in the map.
The hyperfine fitting results can be seriously affected by anomalies. 
For example, the irregularly high intensity of the isolated hyperfine component in
points  4, 5, and 6 in Figure \ref{fig:DCE185} may result in the high optical depth in the best-fit.
In addition, since the line profiles do not saturate even for the strong component, the high optical depth from 
the best-fit is likely affected by hyperfine anomalies.
If we take the optical depth of 0.5--0.7 as an upper limit (1.5--2.1 for 101--012 component), the optically thin 
assumption will result in an underestimation of the column density by a factor of $\sim 1.9-2.4$, which is much 
smaller than the degree of N$_2$H$^+$ depletion toward the center (see section \ref{sec:abundance}).
Therefore, we suggest that assuming optically thin emission is reasonable for calculating the column density with 
the 101--012 component.

\section{Uncertainty of the peak radius of N$_2$H$^+$ abundance, $R_{\rm N_2H^+,peak}$}
\label{sec:power-law_index}
\begin{figure*}
\includegraphics[width=0.95\textwidth]{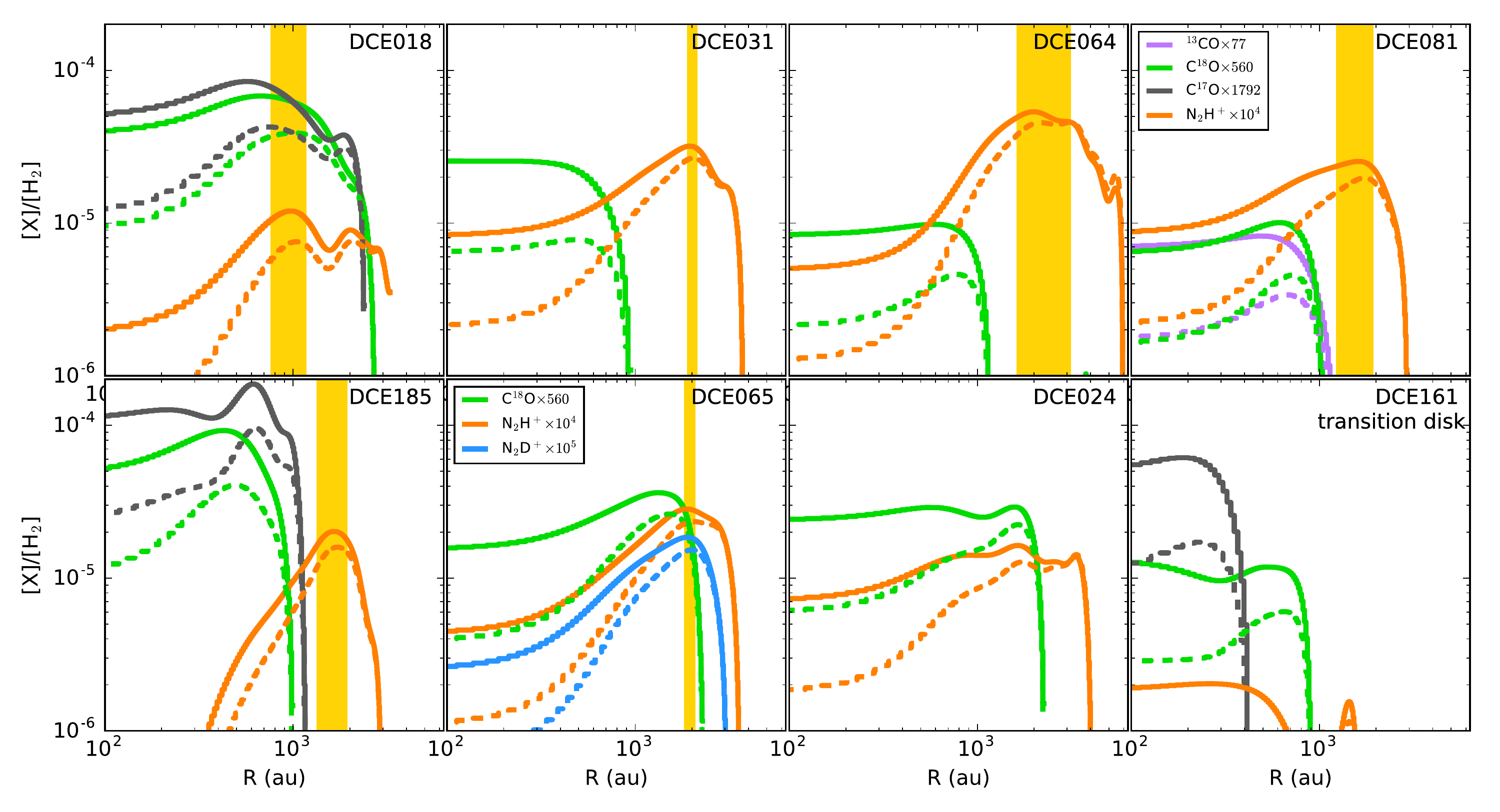}
\caption{ Same as Figure \ref{fig:abundance} but with the density profiles that 
takes the power-law indices of $-1.5$ (solid lines) and $-2.0$ (dashed lines).}
\label{fig:abundance_a}
\end{figure*}

Since the N$_2$H$^+$ abundance profiles are obtained from the ratio between the N$_2$H$^+$ 
and H$_2$ column densities, the uncertainty in $R_{\rm N_2H^+,peak}$ primarily comes from 
(1) the modeled H$_2$ density profiles and (2) differences between the two sides of the cut 
used to derive the abundance profile. 
Thus, we measure the peak radius in the abundance profile on each side of the cut separately. 
For each side, the density structure is assumed to be either 
(1) a broken power-law (Figure \ref{fig:abundance}) in section \ref{sec:abundance}, 
or (2) a single power-law with $p=-1.5$, or 
(3) a single power-law with $p=-2.0$ (Figure \ref{fig:abundance_a}). 
In total, we get six measurements for each source. 
Table \ref{tab:result} lists the range of derived radii, and we take this range as the uncertainty in the measurement.

\label{sec:a}

\end{document}